\documentclass[12pt,a4paper]{article}
\usepackage[utf8]{inputenc}
\usepackage[english]{babel}
\usepackage[margin=1in]{geometry}
\usepackage{parskip}
\usepackage{pdflscape}
\usepackage{listings}
\renewcommand{\baselinestretch}{1.25} 
\usepackage{graphicx} % Required for inserting images
\usepackage{float}
\usepackage{array}
        \newcolumntype{P}[1]{>{\raggedright\arraybackslash}p{#1}}

\usepackage{hyperref}
\usepackage{amsmath,amssymb,dsfont}

\usepackage{multirow}
\usepackage[table]{xcolor}
\usepackage{booktabs}
\usepackage[authoryear]{natbib}

\usepackage[normalem]{ulem}
\useunder{\uline}{\ul}{}

\usepackage{amssymb}

\usepackage{tablefootnote}

\usepackage{tikz}
\usetikzlibrary{arrows.meta}

\usepackage{tcolorbox}
\newtcolorbox{mybox}[1]{colback=black!5!white,colframe=white!50!black,fonttitle=\bfseries,title=#1}

\DeclareMathOperator*{\argmax}{arg\,max}

\newtheorem{definition}{Definition}

\usepackage{adjustbox}

\title{Digital Forgetting in Large Language Models: A Survey of Unlearning Methods}
\author{Alberto Blanco-Justicia$^1$, Najeeb Jebreel$^1$,\\ Benet Manzanares$^1$, David S\'anchez$^1$$^*$, Josep Domingo-Ferrer$^1$$^*$,\\
Guillem Collell$^2$, and Kuan Eeik Tan$^2$\vspace{0.5cm}\\
$^1$ Universitat Rovira i Virgili\\
Department of Computer Engineering and Mathematics\\
CYBERCAT-Center for Cybersecurity Research of Catalonia\\
Av. Pa\"{\i}sos Catalans 26, 43007 Tarragona, Catalonia.\vspace{0.2cm}\\
$^2$ Huawei Technologies Finland Research Center\\
It\"amerenkatu 9, Helsinki, FI-00180 Finland.\vspace{0.2cm}\\
$^*$ Corresponding authors: \{david.sanchez, josep.domingo\}@urv.cat}
\date{\today}

\begin{document}

\maketitle
\newpage
\tableofcontents
\newpage

\section{Introduction}

Large language models have become the state of the art in most if not all 
natural language processing (NLP) and natural language understanding (NLU) tasks.
Since the publication of the transformer architecture by \cite{vaswani2017attention},
several authors have made use of this architecture, or variations of it, to
tackle tasks such as translation, summarization, question answering, sentiment
analysis, or text generation.
Since the announcement and publication of ChatGPT by OpenAI in November 2022,
which brought the LLMs capabilities to a broad audience, several issues have
been raised, mainly concerned with the {\em alignment} of such models to 
societal values and the rule of law\footnote{While this document is dedicated
to LLMs, similar issues have been raised in regard to all generative ML models,
such as image or voice generation.}.
Such concerns include the impact of these models on the labor market,
on the right to privacy of individuals, on copyright laws, on the furthering
of biases and discrimination, and on the potential generation of harmful content,
including content that could be used to damage people.

One proposed solution to these issues is that of digital forgetting. The objective
%JOSEP. Changed "flawed" to "model with undesirable knowledge or behavior", because flawed seems to indicate a model that has some mistake.
of digital forgetting is, given a model with undesirable knowledge or behavior, obtain a new model where the
detected issues are no longer present. However, effective digital forgetting 
mechanisms have to fulfill potentially conflicting requirements: the effectiveness
of forgetting, that is how well the new model has forgotten the undesired knowledge/behavior 
(either with formal guarantees or through empirical evaluation);  the retained 
%JOSEP. added on the desirable tasks
performance of the model on the desirable tasks; and the timeliness and scalability of the forgetting procedure.

This document is organized as follows.
Section~\ref{sec:llm} provides a background on LLMs, including their components,
the types of LLMs, and their usual training pipeline.
Section~\ref{sec:bgforgetting} describes the motivations, types, and desired
properties of digital forgetting.
Section~\ref{sec:approaches} introduces the approaches to digital forgetting in LLMs, among which unlearning methodologies stand out as the state of the art.
Section~\ref{sec:ul_survey} provides a detailed taxonomy of machine unlearning methods for LLMs, and surveys and compares current approaches.
Section~\ref{section:evaluation} details datasets, models and metrics
%JOSEP. Changed timeliness to runtime. According to the dictionary, "timeliness" means "the fact or quality of being done or occurring at a favorable or useful time" (vindria a ser com la qualitat de ser oportú).
used for the evaluation of forgetting, retaining and runtime.
Section~\ref{sec:challenges} discusses challenges in the area.
Finally, we provide some concluding remarks in Section~\ref{sec:conclusions}.

%David: TODO for everybody: properly use cite and citep according to whether the cited work are part of the sentence or not.
%Alberto: commented out for the moment
% \subsection{Contributions and differences with other surveys}

% Not needed for Huawei deliverable, but needed for submitting our survey. At least discuss:

% \begin{itemize}
%     \item Knowledge Unlearning for LLMs: Tasks, Methods, and Challenges \cite{Si2023survey}
%     \item Right to be Forgotten in the Era of Large Language Models: Implications, Challenges, and Solutions \cite{zhang2023right}
%     \item Mention other more general surveys on unlearning in standard ML models \cite{qu2023learn,xu2023machine,nguyen2022survey,shaik2023exploring} and comment that unlearning/forgetting on LLMs is a different enough task to justify dedicated techniques and, therefore, a dedicated survey
%     \item W.r.t. other surverys on unlearnin in LLM, quantify the amount of new works surveyed (6 months period), and the more general focus on forgetting.
% \end{itemize}

%Alberto
\section{Background on large language models}
\label{sec:llm}

%\subsection{Large language models}
%
%Introduce the peculiarities, training workflow, and types of LLMs.

(Large) Language models (LLM) are statistical models that assign probabilities to sequences of words.
These models are currently the state of the art in many natural language processing (NLP)
and understanding (NLU) tasks, such as summarization, translation, question answering, sentiment
analysis, text generation, or chatbots, among many other applications.

The first and most popular transformer architecture was presented by Google engineers
\citeauthor{vaswani2017attention} in \citeyear{vaswani2017attention}.
It was introduced as an encoder-decoder architecture for language translation, but it has
since become the main building block for the current (generative) LLM ecosystem.

%David: could be omitted
Such encoder-decoder architectures were first proposed by \cite{cho2014learning}, using recurrent
%JOSEP. Added meaning of LSTM
neural networks (RNNs). RNNs were substituted by long short-term memories (LSTMs) in seq2seq by \cite{sutskever2014sequence}.
The main limitation of RNN/LSTM-based language models was that these models struggled to analyze or
maintain long-term relationships between words in sequences.
\cite{bahdanau2014neural} added an initial differentiable attention mechanism to the encoder-decoder
RNN architecture.
Finally, in \textit{Attention is all you need} \citep{vaswani2017attention}, RNNs were dropped
in favor of using only the attention mechanism, which gave rise to the transformer model.
Since the introduction of the transformer model in 2017, it has become the main architecture
for (large) language models.
%end

\subsection{Components of the transformer architecture}

\begin{figure}[ht!]
        \small
        \renewcommand{\baselinestretch}{1} 
        \centering
        \scalebox{.80}{\input{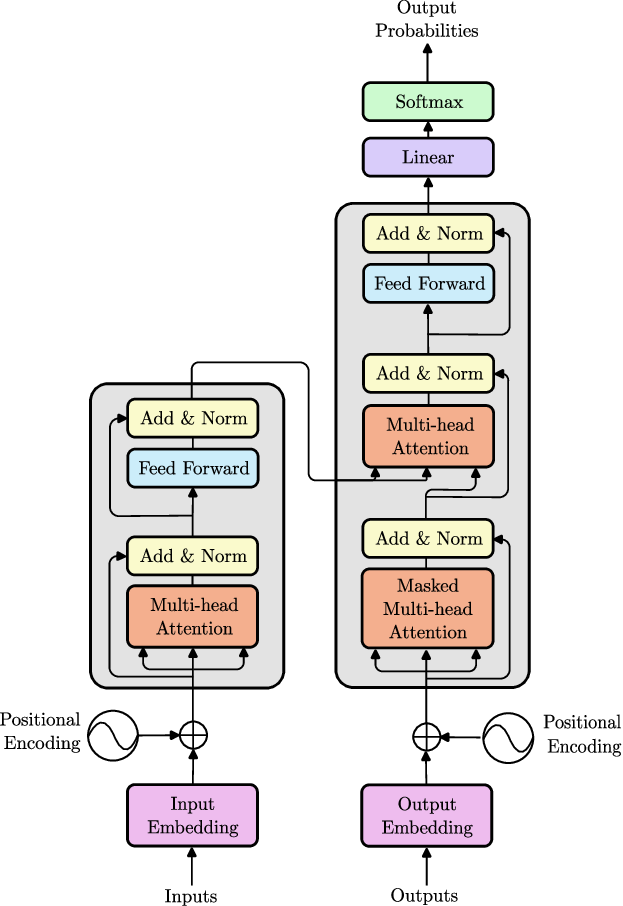}}
        %\includegraphics[scale=.8]{figs/transformer.eps}
%JOSEP. Rewritten caption.
        \caption{The transformer architecture. Note that, since \cite{xiong2020layer}, layer normalization is typically applied before the attention and the feed-forward layers instead of after addition. Image source extracted from \url{https://github.com/negrinho/sane_tikz/blob/master/examples/transformer.tex} under the MIT licence.}
        \label{fig:transformer}
    \end{figure}

The original transformer model, shown in Figure~\ref{fig:transformer},
%JOSEP. I move footnote to caption
%\footnote{Image source extracted from \url{https://github.com/negrinho/sane_tikz/blob/master/examples/transformer.tex} under the MIT licence.}
consisted of an encoder-decoder architecture. 
%JOSEP. I rewrite slightly.
The first components are embedding layers which take in tokens
from the input and output sequences. The resulting embeddings are then passed to the encoder
and decoder components, respectively.
The encoder consists of a number of blocks, each one consisting of two subcomponents: a self-attention layer followed by a feed-forward network. A residual skip connection
adds the output of each subcomponent to its input. In the original transformer
architecture, the outputs of the additions were normalized.
Since the work by \cite{xiong2020layer}, normalization is applied before each subcomponent.
The decoder is similar in structure, but includes an additional self-attention layer
that attends to the outputs of the encoder. 
The first of the self-attention layers in each block of the decoder is 
masked, that is, it only attends to previous tokens (it is causal or autoregressive).
The output of the final decoder block is passed to a dense layer with a softmax output.
Since self-attention is order-independent, position encodings are provided (added) to
the embeddings of the input tokens.

%David: could be omitted
%Alberto: the reasoning for this level of detail is that some of the unlearning
%mechanisms and privacy attacks make explicit references to some of the components,
%I agree that this might be too detailed, and for a survey we could just refer the
%readers to the original papers.
%JOSEP. Added justification of the detail
We describe below the components of the transformer architecture. Details on such components will be needed later when describing some unlearning mechanisms and privacy attacks.

\textbf{Tokenization}. While not exclusive to the transformer model, tokenizers are a central component of most LLMs. LLMs have a limitation on the amount of different words they can deal with. The size of the embedding and the output layers directly depends 
%JOSEP. slightly rewritten.
on the number of different words of the vocabulary in use. In an attempt to minimize the size of such model components (and therefore the memory requirements, training, and inference times), natural language is  \textit{tokenized} in single words or fragments of words. For example, if we have the vocabulary \{quick, quicker, quickest, fast, faster, fastest\}, consisting of $6$ different words, a possible tokenization would be \{quick, fast, er, est\}, of size $4$. Combining the tokens allows us to build the same words as in the original vocabulary, but fewer different tokens are required. The objective of a tokenizer is to cover as much as possible of the target language vocabulary while minimizing the number of tokens. In the following, we will use the terms \textit{words} and \textit{tokens} interchangeably. %SentencePiece
    
\textbf{Embeddings}~\citep{mikolov2013efficient}.
Word embeddings are numerical vector representations of words or tokens. These representations allow for mathematical operations to be applied to words, including distance computations, additions, and subtractions. A notable example of these operations is $vector(King) - vector(Man) + vector(Woman)$, which results in a vector representation close to $vector(Queen)$. An advantage of word embeddings is that they can be trained on general text data, independent of the task, and then be exported to be used as a component in other NLP tasks.
% Common techniques to train word embeddings include the Continuous Bag of Words (CBOW) and Skip-gram models which, respectively, attempt to predict a word within a sequence of words, or attempt to predict sequences of words surrounding a given word.

\textbf{Positional encoding}.
The self-attention mechanism is insensitive to ordering. Thus, explicit order information has to be fed to the transformer. The original architecture used sinusoidal encoding, but relative positions are normally used nowadays.
The position information is added to the vector embeddings before being fed to the attention layers.

\textbf{Multi-head attention}. The attention mechanism enables models to detect the influence or dependence of words or tokens within a sequence (the context) even if the words are not nearby, which solves the main limitation of RNNs and LSTMs (for example, the connection of pronouns to the nouns they refer to in different sentences).
Attention is computed as 
%JOSEP. Slightly edited formula
$$Attn=softmax\left(\frac{\mathbf{Q}\cdot\mathbf{K}^\top}{\sqrt{d_k}}\right)\mathbf{V},$$
where $\mathbf{Q}$, $\mathbf{K}$, and $\mathbf{V}$ are the input sequences multiplied by trainable weight matrices (resp. $W^Q$, $W^K$, and $W^V$), and $d_k$ is the dimension of the weight matrices.
In autoregressive models (like decoder-only transformers), masked self-attention is used instead of plain self-attention. The mask $M$ is a 
%JOSEP. I suppose you mean -infty rather than -inf
triangular matrix where the upper triangle is set to $-\infty$ and the lower triangle is set to $0$. 
%JOSEP. Changed the formula M + \mathbf{QK}^\top to \mathbf{QK}^\top$
$M$ is added to the product $\mathbf{QK}^\top$ to cancel out any influence of \textit{future} tokens on the current token. Note that autoregressive models can only be influenced by past tokens.
The attention mechanism can be easily parallelized (and hence the term multi-head attention). The attention operation is applied to several (depending on a set hyperparameter) input sequences, and the results are concatenated before being fed to the following layers.

\textbf{Feed-forward network}. After attention, a feed-forward network consisting of two linear layers is applied. These two layers do not typically include biases.
%end

\subsection{Types of LLMs}

While the original transformer architecture consisted of an encoder-decoder architecture, variations of this have been introduced in subsequent works. 
Namely, encoder-only architectures such as BERT \citep{devlin2018bert}, decoder-only architectures such as GPT \citep{radford2018improving} and LLaMa \citep{touvron2023llama1}, and encoder-decoder architectures, such as T5 \citep{raffel2020exploring}.
Each of these architectures has been used to tackle different NLU and NLP tasks, and is trained with different pre-training objectives, all of them self-supervised.

%David: could be omitted for the survey
\textbf{Encoder-only models}, such as BERT \citep{devlin2018bert}, use only the encoder blocks of the transformer architecture, which include the full self-attention layers. Thus, these models can access all
words in the given input. 
Encoder models are often used for tasks that require the understanding of full sentences, such as sentence or word classification, or extractive question answering.
The pre-training of these models usually consists of masked language modeling (MLM) and/or next-sentence prediction (NSP). 
% In MLM, a fraction of the training data (15\% for BERT) is masked using a special token and fed to the network. The training objective is to reconstruct the original --unmasked-- sequences. 
% In NSP tasks, pairs of sentences are input to the network, with a special separation token inserted between the sentences. The model is trained to predict whether a sentence naturally follows the previous one.
    
    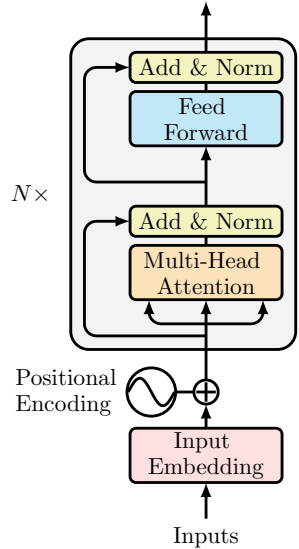
\begin{figure}[ht!]
        \small
        \renewcommand{\baselinestretch}{1} 
        \centering
        \scalebox{.80}{\begin{tikzpicture}
\definecolor{emb_color}{RGB}{252,224,225}
\definecolor{multi_head_attention_color}{RGB}{252,226,187}
\definecolor{add_norm_color}{RGB}{242,243,193}
\definecolor{ff_color}{RGB}{194,232,247}
\definecolor{softmax_color}{RGB}{203,231,207}
\definecolor{linear_color}{RGB}{220,223,240}
\definecolor{gray_bbox_color}{RGB}{243,243,244}

\draw[fill=gray_bbox_color, line width=0.046875cm, rounded corners=0.300000cm] (-0.975000, 6.455000) -- (2.725000, 6.455000) -- (2.725000, 1.305000) -- (-0.975000, 1.305000) -- cycle;
\draw[line width=0.046875cm, fill=emb_color, rounded corners=0.100000cm] (0.000000, 0.000000) -- (2.500000, 0.000000) -- (2.500000, -0.900000) -- (0.000000, -0.900000) -- cycle;
\node[text width=2.500000cm, align=center] at (1.250000,-0.450000) {Input \vspace{-0.05cm} \linebreak Embedding};
\draw[line width=0.046875cm, fill=add_norm_color, rounded corners=0.100000cm] (0.000000, 3.680000) -- (2.500000, 3.680000) -- (2.500000, 3.180000) -- (0.000000, 3.180000) -- cycle;
\node[text width=2.500000cm, align=center] at (1.250000,3.430000) {Add \& Norm};
\draw[line width=0.046875cm, fill=multi_head_attention_color, rounded corners=0.100000cm] (0.000000, 3.030000) -- (2.500000, 3.030000) -- (2.500000, 2.130000) -- (0.000000, 2.130000) -- cycle;
\node[text width=2.500000cm, align=center] at (1.250000,2.580000) {Multi-Head \vspace{-0.05cm} \linebreak Attention};
\draw[line width=0.046875cm] (1.250000, 3.030000) -- (1.250000, 3.180000);
\draw[line width=0.046875cm, fill=add_norm_color, rounded corners=0.100000cm] (0.000000, 6.230000) -- (2.500000, 6.230000) -- (2.500000, 5.730000) -- (0.000000, 5.730000) -- cycle;
\node[text width=2.500000cm, align=center] at (1.250000,5.980000) {Add \& Norm};
\draw[line width=0.046875cm, fill=ff_color, rounded corners=0.100000cm] (0.000000, 5.580000) -- (2.500000, 5.580000) -- (2.500000, 4.680000) -- (0.000000, 4.680000) -- cycle;
\node[text width=2.500000cm, align=center] at (1.250000,5.130000) {Feed \vspace{-0.05cm} \linebreak Forward};
\draw[line width=0.046875cm] (1.250000, 5.580000) -- (1.250000, 5.730000);
\draw[line width=0.046875cm] (1.250000, 0.600000) circle (0.200000);
\draw[line width=0.046875cm] (1.410000, 0.600000) -- (1.090000, 0.600000);
\draw[line width=0.046875cm] (1.250000, 0.760000) -- (1.250000, 0.440000);
\draw[line width=0.046875cm] (0.350000, 0.600000) circle (0.400000);
\draw[line width=0.046875cm] (-0.030000, 0.600000) -- (-0.014490, 0.629156) -- (0.001020, 0.657833) -- (0.016531, 0.685561) -- (0.032041, 0.711884) -- (0.047551, 0.736369) -- (0.063061, 0.758616) -- (0.078571, 0.778258) -- (0.094082, 0.794973) -- (0.109592, 0.808486) -- (0.125102, 0.818576) -- (0.140612, 0.825077) -- (0.156122, 0.827883) -- (0.171633, 0.826946) -- (0.187143, 0.822284) -- (0.202653, 0.813971) -- (0.218163, 0.802145) -- (0.233673, 0.786999) -- (0.249184, 0.768783) -- (0.264694, 0.747796) -- (0.280204, 0.724382) -- (0.295714, 0.698925) -- (0.311224, 0.671845) -- (0.326735, 0.643584) -- (0.342245, 0.614608) -- (0.357755, 0.585392) -- (0.373265, 0.556416) -- (0.388776, 0.528155) -- (0.404286, 0.501075) -- (0.419796, 0.475618) -- (0.435306, 0.452204) -- (0.450816, 0.431217) -- (0.466327, 0.413001) -- (0.481837, 0.397855) -- (0.497347, 0.386029) -- (0.512857, 0.377716) -- (0.528367, 0.373054) -- (0.543878, 0.372117) -- (0.559388, 0.374923) -- (0.574898, 0.381424) -- (0.590408, 0.391514) -- (0.605918, 0.405027) -- (0.621429, 0.421742) -- (0.636939, 0.441384) -- (0.652449, 0.463631) -- (0.667959, 0.488116) -- (0.683469, 0.514439) -- (0.698980, 0.542167) -- (0.714490, 0.570844) -- (0.730000, 0.600000);
\draw[line width=0.046875cm, -latex] (1.250000, 3.680000) -- (1.250000, 4.680000);
\draw[line width=0.046875cm, -latex] (1.250000, 0.000000) -- (1.250000, 0.400000);
\draw[line width=0.046875cm, -latex] (1.250000, 0.800000) -- (1.250000, 2.130000);
\draw[line width=0.046875cm] (0.750000, 0.600000) -- (1.050000, 0.600000);
\draw[-latex, line width=0.046875cm, rounded corners=0.200000cm] (1.250000, 4.080000) -- (-0.750000, 4.080000) -- (-0.750000, 5.980000) -- (0.000000, 5.980000);
\draw[-latex, line width=0.046875cm, rounded corners=0.200000cm] (1.250000, 1.530000) -- (-0.750000, 1.530000) -- (-0.750000, 3.430000) -- (0.000000, 3.430000);
\draw[-latex, line width=0.046875cm, rounded corners=0.200000cm] (1.250000, 1.730000) -- (0.312500, 1.730000) -- (0.312500, 2.130000);
\draw[-latex, line width=0.046875cm, rounded corners=0.200000cm] (1.250000, 1.730000) -- (2.187500, 1.730000) -- (2.187500, 2.130000);
\draw[line width=0.046875cm, -latex] (1.250000, -1.500000) -- (1.250000, -0.900000);
\node[text width=2.500000cm, anchor=north, align=center] at (1.250000,-1.500000) {Inputs};
\node[anchor=east] at (-1.175000,3.880000) {$N\times$};
\node[text width=2.000000cm, anchor=east] at (0.250000,0.600000) {Positional \vspace{-0.05cm} \linebreak Encoding};

\draw[line width=0.046875cm, -latex] (1.250000, 6.230000) -- (1.250000, 7.080000);

% \draw[line width=0.046875cm, fill=linear_color, rounded corners=0.100000cm] (0.000000, 7.580000) -- (2.500000, 7.580000) -- (2.500000, 7.080000) -- (0.000000, 7.080000) -- cycle;
% \node[text width=2.500000cm, align=center] at (1.250000,7.330000) {Linear};

% \draw[line width=0.046875cm, fill=softmax_color, rounded corners=0.100000cm] (0.000000, 8.680000) -- (2.500000, 8.680000) -- (2.500000, 8.180000) -- (0.000000, 8.180000) -- cycle;
% \node[text width=2.500000cm, align=center] at (1.250000,8.430000) {Softmax};
% \draw[line width=0.046875cm, -latex] (1.250000, 7.580000) -- (1.250000, 8.180000);
% \draw[line width=0.046875cm, -latex] (1.250000, 8.680000) -- (1.250000, 9.300000);
% \node[text width=2.500000cm, align=center] at (1.250000,9.50000) {Class};

\end{tikzpicture}}
        \caption{Encoder-only transformer}
        \label{fig:encoder}
    \end{figure}

\textbf{Decoder-only models}, such as GPT \citep{radford2018improving}, only make use of the (non-conditioned) decoder blocks of the transformer model, which contain the masked attention layers. Therefore, these models can only attend to past tokens. These models are often called auto-regressive models.
Decoder models are used in generative tasks and are usually pre-trained with next-word prediction (NWP) tasks.
% In NWP tasks, training data is constructed by taking sequences of tokens and using the last token of the sequence as the classification label.
     
    \begin{figure}[ht!]
        \small
        \renewcommand{\baselinestretch}{1} 
        \centering
        \scalebox{.80}{\begin{tikzpicture}
\definecolor{emb_color}{RGB}{252,224,225}
\definecolor{multi_head_attention_color}{RGB}{252,226,187}
\definecolor{add_norm_color}{RGB}{242,243,193}
\definecolor{ff_color}{RGB}{194,232,247}
\definecolor{softmax_color}{RGB}{203,231,207}
\definecolor{linear_color}{RGB}{220,223,240}
\definecolor{gray_bbox_color}{RGB}{243,243,244}

\draw[fill=gray_bbox_color, line width=0.046875cm, rounded corners=0.300000cm] (3.775000, 6.855000) -- (7.475000, 6.855000) -- (7.475000, 1.305000) -- (3.775000, 1.305000) -- cycle;
\draw[line width=0.046875cm, fill=emb_color, rounded corners=0.100000cm] (4.000000, 0.000000) -- (6.500000, 0.000000) -- (6.500000, -0.900000) -- (4.000000, -0.900000) -- cycle;
\node[text width=2.500000cm, align=center] at (5.250000,-0.450000) {Embedding};

\draw[line width=0.046875cm, fill=add_norm_color, rounded corners=0.100000cm] (4.000000, 4.080000) -- (6.500000, 4.080000) -- (6.500000, 3.580000) -- (4.000000, 3.580000) -- cycle;
\node[text width=2.500000cm, align=center] at (5.250000,3.830000) {Add \& Norm};
\draw[line width=0.046875cm, fill=multi_head_attention_color, rounded corners=0.100000cm] (4.000000, 3.430000) -- (6.500000, 3.430000) -- (6.500000, 2.130000) -- (4.000000, 2.130000) -- cycle;
\node[text width=2.500000cm, align=center] at (5.250000,2.780000) {Masked \vspace{-0.05cm} \linebreak Multi-Head \vspace{-0.05cm} \linebreak Attention};
\draw[line width=0.046875cm] (5.250000, 3.430000) -- (5.250000, 3.580000);

\draw[line width=0.046875cm, fill=add_norm_color, rounded corners=0.100000cm] (4.000000, 6.630000) -- (6.500000, 6.630000) -- (6.500000, 6.130000) -- (4.000000, 6.130000) -- cycle;
\node[text width=2.500000cm, align=center] at (5.250000,6.380000) {Add \& Norm};

\draw[line width=0.046875cm] (5.250000, 5.980000) -- (5.250000, 6.130000);

\draw[line width=0.046875cm, fill=ff_color, rounded corners=0.100000cm] (4.000000, 5.980000) -- (6.500000, 5.980000) -- (6.500000, 5.080000) -- (4.000000, 5.080000) -- cycle;
\node[text width=2.500000cm, align=center] at (5.250000,5.530000) {Feed \vspace{-0.05cm} \linebreak Forward};

\draw[line width=0.046875cm] (5.250000, 8.530000) -- (5.250000, 8.680000);
\draw[line width=0.046875cm, fill=linear_color, rounded corners=0.100000cm] (4.000000, 7.730000) -- (6.500000, 7.730000) -- (6.500000, 7.230000) -- (4.000000, 7.230000) -- cycle;
\node[text width=2.500000cm, align=center] at (5.250000,7.480000) {Linear};
\draw[line width=0.046875cm, fill=softmax_color, rounded corners=0.100000cm] (4.000000, 8.830000) -- (6.500000, 8.830000) -- (6.500000, 8.330000) -- (4.000000, 8.330000) -- cycle;
\node[text width=2.500000cm, align=center] at (5.250000,8.580000) {Softmax};
\draw[line width=0.046875cm] (5.250000, 0.600000) circle (0.200000);
\draw[line width=0.046875cm] (5.410000, 0.600000) -- (5.090000, 0.600000);
\draw[line width=0.046875cm] (5.250000, 0.760000) -- (5.250000, 0.440000);
\draw[line width=0.046875cm] (6.150000, 0.600000) circle (0.400000);
\draw[line width=0.046875cm] (5.770000, 0.600000) -- (5.785510, 0.629156) -- (5.801020, 0.657833) -- (5.816531, 0.685561) -- (5.832041, 0.711884) -- (5.847551, 0.736369) -- (5.863061, 0.758616) -- (5.878571, 0.778258) -- (5.894082, 0.794973) -- (5.909592, 0.808486) -- (5.925102, 0.818576) -- (5.940612, 0.825077) -- (5.956122, 0.827883) -- (5.971633, 0.826946) -- (5.987143, 0.822284) -- (6.002653, 0.813971) -- (6.018163, 0.802145) -- (6.033673, 0.786999) -- (6.049184, 0.768783) -- (6.064694, 0.747796) -- (6.080204, 0.724382) -- (6.095714, 0.698925) -- (6.111224, 0.671845) -- (6.126735, 0.643584) -- (6.142245, 0.614608) -- (6.157755, 0.585392) -- (6.173265, 0.556416) -- (6.188776, 0.528155) -- (6.204286, 0.501075) -- (6.219796, 0.475618) -- (6.235306, 0.452204) -- (6.250816, 0.431217) -- (6.266327, 0.413001) -- (6.281837, 0.397855) -- (6.297347, 0.386029) -- (6.312857, 0.377716) -- (6.328367, 0.373054) -- (6.343878, 0.372117) -- (6.359388, 0.374923) -- (6.374898, 0.381424) -- (6.390408, 0.391514) -- (6.405918, 0.405027) -- (6.421429, 0.421742) -- (6.436939, 0.441384) -- (6.452449, 0.463631) -- (6.467959, 0.488116) -- (6.483469, 0.514439) -- (6.498980, 0.542167) -- (6.514490, 0.570844) -- (6.530000, 0.600000);

\draw[line width=0.046875cm, -latex] (5.250000, 4.080000) -- (5.250000, 5.080000);
\draw[line width=0.046875cm, -latex] (5.250000, 6.630000) -- (5.250000, 7.230000);
\draw[line width=0.046875cm, -latex] (5.250000, 7.730000) -- (5.250000, 8.330000);
\draw[line width=0.046875cm, -latex] (5.250000, 0.800000) -- (5.250000, 2.130000);
\draw[line width=0.046875cm, -latex] (5.250000, 0.000000) -- (5.250000, 0.400000);
\draw[line width=0.046875cm] (5.450000, 0.600000) -- (5.750000, 0.600000);
\draw[-latex, line width=0.046875cm, rounded corners=0.200000cm] (5.250000, 1.530000) -- (7.250000, 1.530000) -- (7.250000, 3.830000) -- (6.500000, 3.830000);
\draw[-latex, line width=0.046875cm, rounded corners=0.200000cm] (5.250000, 4.480000) -- (7.250000, 4.480000) -- (7.250000, 6.380000) -- (6.500000, 6.380000);
\draw[-latex, line width=0.046875cm, rounded corners=0.200000cm] (5.250000, 1.730000) -- (4.312500, 1.730000) -- (4.312500, 2.130000);
\draw[-latex, line width=0.046875cm, rounded corners=0.200000cm] (5.250000, 1.730000) -- (6.187500, 1.730000) -- (6.187500, 2.130000);

\draw[line width=0.046875cm, -latex] (5.250000, -1.500000) -- (5.250000, -0.900000);
\draw[line width=0.046875cm, -latex] (5.250000, 8.830000) -- (5.250000, 9.430000);
\node[text width=2.500000cm, anchor=north, align=center] at (5.250000,-1.500000) {Inputs};
\node[text width=2.500000cm, anchor=south, align=center] at (5.250000,9.430000) {Next token \vspace{-0.05cm} \linebreak Probabilities};
\node[anchor=west] at (7.675000,4.080000) {$N\times$};
\node[text width=2.000000cm, anchor=west] at (6.750000,0.600000) {Positional \vspace{-0.05cm} \linebreak Encoding};

\end{tikzpicture}}
        \caption{Decoder-only transformer}
        \label{fig:decoder}
    \end{figure}
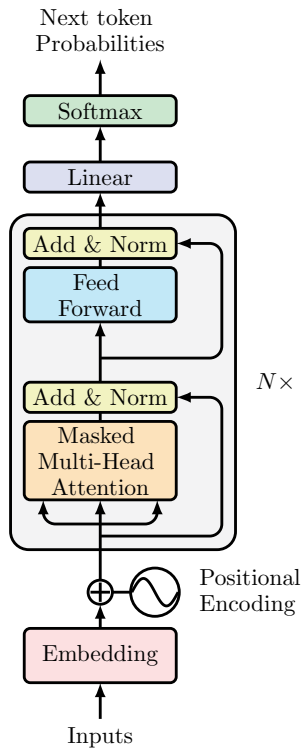

\textbf{Encoder-decoder models} (or sequence-to-sequence models), such as the original transformer and T5 \citep{vaswani2017attention,raffel2020exploring}, use both the encoder and the conditioned decoder blocks of the transformer architecture.
These models are suitable for tasks that generate text conditioned on some other input text, such as translation, summarization, or generative question answering.
The pre-training objective employed by \cite{raffel2020exploring} consists of MLM plus NWP.
% The input sequences are randomly masked with special sentinel tokens. Then, the output sequences consist of the sentinel tokens and their original value. The expected output \textit{class} are the values of the masked tokens.
    
%David: perhaps too descriptive? Pros and cons of types of applications of each type of model could be discussed here to better understand the implications of their designs?
%Alberto: so maybe some text saying that encoder-only is mostly used for classification tasks, decoder-only for text generation, and encoder-decoder mostly for text-to-text tasks, such as translation? At least out of the box: fine-tuning can blur these distinctions.
%Also, the main difference between encoders and decoders is in the attention heads, decoders are masked to only 'see' past tokens, while encoders can focus on the whole input.

\subsection{Training of large language models}

The training of LLMs often occurs in different phases, depending on the objectives.
These phases include a self-supervised pre-training phase, with tasks or training objectives
that depend on the type of LLM being trained; a fine-tuning phase, where labeled data are used
to train specific tasks, such as sentiment analysis, conversational text generation,
or summarization; an additional round of fine-tuning using reinforcement learning from human
feedback (RLHF) in which human annotators help fine-tuning the model by providing feedback
on the model responses; and a final phase, where specific {\em system prompts} can be passed
to the model to guide it and condition its future responses.

We consider of interest to this document any model that has at least undergone a pre-training
phase. These pre-trained models are often called {\em foundation} models. 
In the following, we describe these training phases.

%David: be sure that all acronyms are properly introduced the first time they are mentioned. I've added several ones, but I may have missed some of them.
\textbf{Pre-training}: self-supervised learning of different tasks, with standard cross-entropy losses and stochastic gradient descent (SGD) --Adam-- optimizers. 
%David: probably not needed
We next exemplify some of the most common pre-training tasks given the original sentence \texttt{My cat likes playing with the ball.}, found in the pre-training dataset.

\begin{itemize} 
%JOSEP. The MLM acronym has already been introduced above.
    \item \emph{Masked Language Modeling}. In MLM, random tokens from the input sequences are masked with a special token. Models are trained to predict the missing token. 

    \begin{mybox}{Example of MLM}
    Input: \texttt{My cat <MASK> playing with the ball.}\\
    Label: \texttt{likes}
    \end{mybox}

%JOSEP: The NSP acronym has been introduced above.
    \item \emph{Next Sentence Prediction}. In NSP, the models are trained to predict whether a given sentence logically follows another sentence. The two input sentences are separated with a special token.
    
    \begin{mybox}{Example of NSP}
    \raggedright
    Input: \texttt{My cat likes playing with the ball.<SEP>He likes playing with other toys, too.}\\
    Label: \texttt{IsNextSentence}\\
    Input: \texttt{My cat likes playing with the ball.<SEP>Harry Potter is a wizard.}\\
    Label: \texttt{NotNextSentence}
    \end{mybox}

%JOSEP. The NWP acronym has already been introduced above.
    \item \emph{Next Word Prediction}. In NWP, the model is instructed to return the next word (token) given some text sequence. Single sentences can be used as multiple training inputs, as shown below.

%JOSEP. NTP-> NWP
    \begin{mybox}{Example of NWP}
    Input: \texttt{My}\\
    Label: \texttt{cat}\\
    Input: \texttt{My cat likes playing}\\
    Label: \texttt{with}\\
    Input: \texttt{My cat likes playing with the}\\
    Label: \texttt{ball}
    \end{mybox}
    
    \item \emph{T5 Task}. The T5 transformer is an encoder-decoder architecture, and thus the pre-training is performed in text-to-text tasks, also including MLM.
    \begin{mybox}{Example of T5 pre-training}
    \raggedright
    Input: \texttt{My <X> likes <Y> with the <Z>.}\\
    Output: \texttt{<X> cat <Y> playing <Z> ball.}\\
    Labels: \texttt{cat, playing, ball}
    \end{mybox}
\end{itemize}
%end
    
\textbf{Fine-tuning}: further supervised training conducted on domain-related labeled data to teach new
(downstream) tasks or knowledge domains to a pre-trained model~\citep{touvron2023llama2}.
For example, an encoder-only transformer such as BERT can be fine-tuned to conduct sentiment analysis by
passing the model sentences and sentiment pairs.
Another such example is GitHub Copilot\footnote{GitHub Copilot -- \url{https://github.com/features/copilot}},
where a pre-trained decoder-only model is further trained on code.
Chatbots, such as Llama2-Chat or GPT-3.5-turbo, are fine-tuned on question-answer pairs.
Adapters, including LoRA~\citep{hu2021lora}, can be used to reduce the cost of fine-tuning by
freezing the pre-trained model weights and injecting trainable rank decomposition matrices
into each layer of the transformer architecture, thereby reducing the number of trainable
parameters for downstream tasks.
    
\textbf{Reinforced learning from human feedback (RLHF)} \citep{christiano2017deep}.
%Reinforcement learning, PPO, competing answers, human rating, ELO, adaptation.
In RLHF, models generate different outputs for a given prompt. Then, human annotators rank the quality or appropriateness of the answers produced by the model. These new data, that is, the prompts, the related answers, and the given rankings of answers, are used to train a reward model (as in reinforcement learning) that can then be used to further train the original model given the rewards the model produces for each answer. RLHF has been used to build ChatGPT\footnote{OpenAI, Introducing ChatGPT -- \url{https://openai.com/blog/chatgpt}}, to make the answers of the original model (GPT-3, GPT-3.5, or GPT-4) more conversational and aligned with OpenAI's objectives.
Meta uses RLHF to improve the models usefulness and safety \citep{touvron2023llama2}.
    
\textbf{Prompt engineering} (zero-shot, one-shot, few-shot learning): the behavior of LLMs can be further steered by including appropriate information in the prompts given to the models~\citep{brown2020language,min2022rethinking,wei2023chainofthought}. For example, an LLM can be given some examples of how to solve a problem, with the formulation of the problem and its solution, before asking it the solution of a similar problem (few-shot learning). Likewise, models can be instructed on how to respond or not to respond to specific questions. Production models such as ChatGPT include \textit{secret} instructions inserted by the provider to guide their general operations. These ``in-context learning'' or ``prompt engineering'' approaches, however, can be
sometimes bypassed by other user-provided instructions.

\section{Digital forgetting}
\label{sec:bgforgetting}
%David: add an introductory paragraph defining DF and the structure of the section

As described above, LLMs undergo an initial pre-training phase using text mostly gathered
from the internet (and often uncurated). This opens the possibility for models to be trained
on private, biased, or copyright-protected data, and even to pick up harmful or hateful
behaviors. Additionally, these issues can be further augmented during fine-tuning if the 
labeled data is of poor quality.

This section discusses the motivations for digital forgetting, including legal and ethical issues,
the types of digital forgetting, and the desirable properties of forgetting procedures.

\subsection{Motivations for digital forgetting}
\label{section:motivations}
The need for digital forgetting stems from several ethical principles, national and supranational
regulations, and codes of conduct. In general, regulations relating to the protection of the privacy
of individuals ({\em e.g.}, GDPR) and the protection of intellectual property ({\em e.g.}, copyright
laws) motivate the research, discussion, and implementation of digital forgetting measures in software
systems, including large language models. 

In the following, we categorize and discuss the reasons for implementing digital forgetting in ML. 

%David: probably not needed, even though some instances could be mentioned in the next subsections
%\begin{itemize}
%    \item Universal Declaration of Human Rights and EU Charter of Fundamental Rights
%    \item GDPR~\cite{GDPR}
%    \begin{itemize}
%        \item Some introduction to GDPR: actors, rights, and obligations.
%        \item Articles 12--23 encode data subject rights, including the Right of Rectification, and the Right of Erasure (\textit{right to be forgotten}).
%        Article 17: Right to Erasure:
%        ``The data subject shall have the right to obtain from the controller the erasure of personal data concerning him or her without undue delay and the controller shall have the obligation to erase personal data without undue delay where one of the following grounds applies.'' The data is no longer necessary, the subject withdraws consent, the subject objects to processing, or the data have been unlawfully processed.
%    \end{itemize}
%    \item Copyright law and generative models (positions by different countries -- Hiroshima AI Process)
%    \begin{itemize}
%        \item WIPO: copyright law does not protect ideas, but the form they are expressed in.
%        \item Initial stances by US, EU, Japan
%        \item G7 Hiroshima Process on AI~\cite{OECDHiroshima,G7Hiroshima}
%        \item EU AI Act: list/inform of copyrighted material used for training.
%    \end{itemize}
%    \item Bad behaviors -- Responsible / EU Guidelines for Trustworthy AI~\cite{trustworthyAI}
%    \item EU AI Act
%\end{itemize}
%end

%Reasons to forget (See \cite{nguyen2022survey}):

\subsubsection{Privacy}
\label{section:motivations-privacy}

ML models, including LLMs, are trained on vast amounts of data, often obtained from open sources on the web. 
Misconfigured services could include private data, either personal or internal data from organizations,
which can be indexed by search engines and freely accessed. These data, which are unintended to be public,
may end up in the training datasets used for LLM pre-training.
Additionally, private data may be used to fine-tune pre-trained models to teach them new downstream tasks.
For example, a hospital could fine-tune a pre-trained model using data from patients to teach the model
how to diagnose new patients.

Machine learning models in general have been shown to memorize and leak data from the training dataset~\citep{shokri2017membership,salem2018ml}.
Outlying data are even more at risk of being memorized and leaked, which in the case of personal
data may lead to privacy risks~\citep{smith2023identifying}. 
One of the most basic forms of attacks on ML models are membership inference attacks (MIA),
in which an attacker attempts to determine whether a given data point is part of the training dataset.
In essence, most MIAs consist of finding out how the models behave when exposed to data from the
training dataset compared to previously unseen data.
This difference in behavior can be measured by the differences in the loss, the classification 
confidence, or other metrics. In the case of LLMs, a common approach is to focus on the perplexity metric, 
%JOSEP. Added explanation on perplexity
which measures how certain a model is about a given sequence of
text. 

\cite{yeom2018privacy} established a connection between the vulnerability to membership inference attacks
and overfitting. While this has been confirmed by other works~\citep{surveyDP}, overfitting does not seem to be the only source of vulnerability.
As an example, pre-trained LLMs are seldom overfitted to the training data.
First, because they use vast datasets to train, and second, because they are trained for very
few epochs (sometimes just a single epoch).
However, there have been some attacks that show that LLMs memorize the training data even when not overfitted.

\subsubsection{Copyright protection}
\label{section:motivations-copyright}

%In the copyright content removal scenario, unlearning can efficiently erase copyrighted information from LLMs. 
The copyright protection objective for digital forgetting in LLMs is closely related to privacy, as in both cases we do not want the generated text to include specific information. 
%In generative models, we do not want generated text to include personal information about individuals even if private information was present in the training data.
%Likewise, we do not want the models to output \textit{verbatim} fragments of copyrighted text, which will very likely be included in the training data given how LLMs are trained.
However, there is a key distinction: in privacy protection, we want to avoid disclosing personal information expressed in any way, whereas in copyright protection we specifically focus on verbatim text. 

A model that is capable of answering questions about an individual and whose answers contain sensitive details about such an individual is infringing on their right to privacy. This is not the case with copyright. Copyright laws do not protect facts, but the exact form in which they are expressed. Therefore, a model that can answer questions about some copyrighted work does not necessarily infringe copyright law unless verbatim fragments of the work are generated (and still, the law includes provisions to lawfully quote protected works). Therefore, any information extraction attacks that consider ``exact'' information may be enough to check for compliance regarding copyright 
%JOSEP. Added parenthetical explanations and rewritten.
laws (copyright infringement requires verbatim information to be extracted), but they are not enough to check privacy compliance (privacy can be infringed even if the information extracted about someone has been rephrased).

%David: this and the next subsection may be merged (and developed a little bit further), as malicious/poisoned data can be considered a type of noisy data to be forgotten to improve the model performance
% \subsubsection{Improved model performance}
% Forgetting the influence of noisy, malicious, outlier, and outdated data can enhance LLMs' performance and robustness. \textit{Again Trustworthy AI. More in general, this could refer to the "alignment" requirements. Humans should remain in control.}

% \subsubsection{Security}
% LLMs might be trained on poisoned data which can mislead models. Forgetting can be employed to eradicate the poison from the model. \textit{Trustworthy AI requirement of robustness against attacks}.

%Alberto: merged and renamed to Model robustness
\subsubsection{Model robustness}

The whole LLM training pipeline includes pre-training from public data, fine-tuning with 
public, crowdsourced, or proprietary data, and possibly further fine-tuning with RLHF, which
may also be public, crowdsourced, or closed.
In all these phases of training, there is a possibility to process low-quality information.
Datasets used for pre-training may include false, inconsistent, or outdated information.
Datasets used for fine-tuning may be crowdsourced, which opens the door to malicious actors providing wrong information.
RLHF depends on human annotators, which rank model outputs based not only on usefulness but also on safety. Rogue annotators may provide wrong information.

All of these sources of outdated information, misinformation, outliers, noise, and malicious reporting may influence the learning process and thus produce underperforming models with potentially critical failures.

Forgetting procedures may be used to correct some of these issues.

%%%%%%%%%%%%%%%%%%%%%%%%

%Alberto: changed from Fairness to be more general and include harmful content
\subsubsection{Alignment with human values}
% Forgetting may be necessary to address biases in LLMs. Removing biased data is essential to ensure fairness and prevent discriminatory outcomes.
% One identified source of bias in LLMs is gender bias. It has been identified that the pronouns he and she are strongly correlated with different careers or occupations.
%David: Huawei is especifically interested in the removal of 'harmful' content (e.g. prompting LLM for methods to self-harm, harm others, suggest  violent act, etc.), hate speech, etc.. Should this be included here? In any case, this should be developed in more detail, since it is the main priority for Huawei.
%Alberto: Here's an attempt
LLM pre-training datasets are compiled from diverse, often uncurated sources such as web pages, online encyclopedias, social media posts, and online book libraries. Some of these sources may contain content misaligned with current societal values, including discriminatory language based on gender, race, ethnicity, sexual orientation, or age, often manifesting as stereotypes or prejudices.
Notably, studies have identified strong correlations between pronouns (he/she) and different careers.
Furthermore, these sources may include hateful, violent, or harmful speech associated with discrimination.
Pre-training models on such data may not only continue these biased and harmful behaviors but even amplify them in some cases.

Machine learning models, including LLMs, must not only protect individual
privacy rights, as discussed above, but also adhere to ethical values and principles, taking as a
reference point the Universal Declaration of Human Rights~\citep{UDHR}, and also including laws and
social norms. Alignment with principles such as non-discrimination, fairness, benevolence, and
non-maleficence is of utmost importance. In addition, the EU Guidelines 
%JOSEP. Corrected author name in bibliography.
for Trustworthy AI~\citep{trustworthyAI} require that developed machine learning models be always under human
control and supervision, which means that any deviation from such principles should be addressed
(or addressable). Therefore, alignment with ethical values may prompt requests for model forgetting
procedures.

Forgetting procedures can be employed to identify and eliminate such sources of discriminatory or
harmful behavior, aligning the models with prevailing social norms.

\subsection{Types of digital forgetting}
\label{section:types}

%Benet: Menciono aquí "undesired knowledge" por primera vez pk luego lo uso en Evaluation
We next discuss the types of undesired knowledge that may be subjected to forgetting in LLMs. 

\textbf{General forgetting request.}
%JOSEP. Slightly rewritten paragraph.
The Right to Erasure in Article 17 of the GDPR states that: ``The data subject shall have the right to obtain from the controller the erasure of personal data concerning him or her without undue delay and the controller shall have the obligation to erase personal data without undue delay where one of the following grounds applies.''~\citep{GDPR} The regulation does not specify any specific form or information the data subject has to provide to the data controller to exercise their right. 
Search engines, such as Google Search, provide specific forms to exercise the right to be forgotten. In their forms, Google requests the data subjects to identify, and then provide a list of URLs that contain personal information about them, which specific search queries point to the documents of interest, and the motivation for the erasure\footnote{Google's Right to be Forgotten form -- \url{https://reportcontent.google.com/forms/rtbf}}.
Therefore, the implementation of the right to be forgotten revolves around removing \textit{documents} that are present on the internet (and not published by the data subjects themselves) and contain some personal information about the data subject from search results.
This erasure does not imply \textit{all} information about a data subject has to be deleted from search results, but only that information requested by the data subject.
How this translates to LLMs is a subject to be further analyzed. 

%David: to be changed when the discussion of the attacks is reallocated 
%Alberto: changed ref
%JOSEP. Restructured paragraph and slightly rewritten.
In Section~\ref{sec:mia}, we describe a series of knowledge memorization definitions and attacks
that shed some light on how a forgetting request may be studied and acted on. 
Definition~\ref{def:knowl_extraction} in that section states that some information is extractable from an LLM if there is some \textit{context} of \textit{prompt} for which the LLM returns said information. Definition~\ref{def:extractable_memo} follows similar lines.
Connecting those definitions to forgetting requests as implemented by Google (and other services) amounts to providing a series of prompts for which private data about the data subject are returned. Note that in the case of generative LLMs, the generation is stochastic and depends on a sampling strategy from the generated LLM distributions; therefore, finding appropriate prompts may not be feasible.
Definition~\ref{def:discoverable_memo}, however, mentions prompt-response pairs; hence, testing whether an LLM returns private information about a topic having the specific document that should be deleted should be straightforward (as shown by attacks described in the same Section~\ref{sec:mia}). 

%JOSEP. Slightly rewritten rest of section.
\textbf{Item removal requests.} These want forgetting of one or more specific data points or samples contained within the model. Such requests are straightforward for models dealing with tabular or image data, where data points are precisely defined. In the realm of tabular data, each row within a table is considered a data point, with one of the attributes serving as the class label for classification or regression. Likewise, in computer vision tasks, individual images are the designated data points.

However, the distinction between items becomes less evident in natural language processing (NLP) tasks and LLMs. While one might consider each token in a text dataset as an individual data point, these tokens often lack meaning on their own. Consequently, in NLP data points can consist of entire sentences or even whole documents, as the separation between meaningful units is less clear-cut compared to tabular or image-based data scenarios.
    
\textbf{Feature or concept removal requests.} These want the model to forget all information about a given subject. The information on a subject may be spread across different sentences and documents. An example of such a request is given by \cite{eldan2023harry}, where the authors attempt to make a model forget everything about the Harry Potter books. A similar approach could be followed to comply with privacy requirements, where all information about a data subject is required to be removed from a model.
    
\textbf{Class removal requests.} These consist of removing all information about a whole class from a model. These requests are quite natural for models used to identify people. For example, in facial recognition, each of the classes corresponds to a single individual, and data points in the class are images of said individual. Thus, removing one class amounts to making the model unable to identify a person that the model could previously identify. A trivial approach for such requests would be to zero out the logits corresponding to the appropriate classes during inference. However, 
%this is not the same as forgetting.
%David: because...
this is only feasible in a black-box setting where the service provider controls the inference. 

Class removal requests can make sense for NLP classification tasks, such as sentiment analysis, in which sentences are classified as being positive, negative, or any other range of sentiments. However, for generative models, each class corresponds to a word or token in the vocabulary of the model.
    
\textbf{Task removal requests.} As described in previous sections, LLMs are pre-trained on large text datasets with generic tasks, such as next-word prediction or masked language modeling. After pre-training, models are fine-tuned to teach them different tasks, such as summarization, sentiment analysis, code writing, conversational text generation, etc. Task removal requests attempt to make the fine-tuned LLMs forget one or more of the tasks that the model has been taught to perform.

%David: separated from the previous section and changed the titles
\subsection{Requirements of digital forgetting}
\label{section:requirements}

%JOSEP. Slightly rewritten.
Regardless of the reason or type of forgetting request, we can define a series of general requirements
that ensure the forgetting procedure is carried out correctly and the resulting model still performs
adequately. This is the purpose of this section, but first we
introduce some preliminary definitions.

A dataset $D$ consists of samples $\{x_i,y_i\}_{i=1}^N$, where $N$ is the size of the dataset, $x_i$ are token sequences, and $y_i$ the true labels. Note that in self-supervised training, as is the case for most pre-training tasks in LLMs, the labels $y_i$ do not need to be explicitly defined. For example, 
%JOSEP. NTP -> NLP
in NLP $y_i$ is the token immediately following the sequence $x_i$ in the text. However, during fine-tuning, labels are expected to be explicitly provided. A forget dataset $D_f \subset D$ is the set of samples to be forgotten. The retain set is defined as $D_r = D \setminus D_f$.

A learning algorithm $A(D)$ is a probabilistic algorithm that outputs a model given a training dataset $D$. Due to the probabilistic nature of $A(\cdot)$, we cannot ensure that running the learning algorithm twice on the same dataset will return the same model.
%JOSEP. I delete this, because mathematically this is always true!
%\textit{i.e.}, the equality $A(D)=A(D)$ is not necessarily true.
%JOSEP. Rewritten below.
However, we can define a probability distribution $P(A(D))$ over all possible models returned by the learning algorithm when trained on the same dataset $D$. Additionally, we can define $Dist(\cdot,\cdot)$ as the distance between two probability distributions. An example of such a distance is the Kullback-Leibler (KL) divergence.

A forgetting algorithm $F(D_f, A(D))$ is a (probabilistic) algorithm that returns a model in which the influence of the samples in $D_f$ has been removed from $A(D)$.
%JOSEP. Rewritten rest of paragraph.
Definitions of unlearning by \cite{nguyen2022survey,xu2023machine} include $D$ as a parameter of the unlearning mechanism (as in $F(D, D_f, A(D))$), but access to $D$ is not always feasible.
For example, a service provider that leverages a foundation model such as Llama2 to offer some service
(possibly after some additional fine-tuning) does not have access to the data used by the party that
conducted the pre-training. However, the service provider may still be required by law to fulfill forgetting requests. For the sake of generality, we will abstain from including $D$ as a parameter for forgetting, although it may be used in some procedures. 

When implementing digital forgetting in ML models (and in LLMs in particular)
the following requirements should be taken into consideration.

%David: I think this itemize would be of great interest for Huawei, but as it is now it is quite sketchy. Could this be developed more?
\textbf{Forgetting guarantees.} An essential requirement for any forgetting procedure, whether in the conventional scenario of search engines or with large language models, lies in the ability to demonstrate the fulfillment of a forgetting request, particularly when 
%JOSEP. Rewritten rest of paragraph.
such fulfillment is a legal obligation. 
A forgetting guarantee serves as a theoretical proof, offering assurance that the content associated with a forgetting request has been forgotten, accompanied by a level of certainty. This ensures a transparent and accountable process in meeting legal requirements and reinforces the credibility of the forgetting mechanism.
\cite{nguyen2022survey} refer to two levels of guarantees, \textit{exact} and \textit{approximate}.
\cite{xu2023machine} further expand approximate forgetting into \textit{strong} and \textit{weak} forgetting, where strong forgetting is equivalent to the definition of approximate forgetting by \cite{nguyen2022survey}, and weak forgetting only applies to the logits of the model, which might be a sufficient guarantee for grey-box and black-box settings.
However, most of the literature on unlearning mechanisms provides no provable guarantees, relying instead on empirical evaluation or auditing.
\begin{itemize}
    \item {\em Exact forgetting.} A forgetting algorithm $F(D_f, A(D))$ provides an \textit{exact} forgetting guarantee 
    if $$Dist(P(F(D_f, A(D))),P(A(D \setminus D_f))=0.$$
    From this definition, we can conclude that retraining from scratch on the retain
    set $D_r$ is a straightforward approach to exact unlearning, since
    %JOSEP. Added missing final parentheses.
    $P(A(D_r))=P(A(D \setminus D_f))$

    \item {\em Approximate forgetting.} A forgetting algorithm $F(D_f, A(D))$ provides an \textit{approximate}
    forgetting guarantee, if $$Dist(P(F(D_f, A(D))),P(A(D \setminus D_f)) \leq t,$$ for an acceptable threshold $t$.
    \citeauthor{nguyen2022survey} provide a definition for $\epsilon$-certified forgetting, inspired by differential privacy. Given $\epsilon > 0$ and a sample $z \in D$,
    $$ e^{-\epsilon} \le \frac{Pr(F(z,A(D)))}{Pr(A(D \setminus z))} \le e^\epsilon.$$

    \item {\em No guarantees.} The forgetting guarantees described above may not be attainable in all cases, either because achieving them or computing them is unfeasible. In these cases, we can refer to empirical evaluation or auditing to provide a level of risk
    %JOSEP. Added parenthetical explanation.
    reduction (that is, how much the risk of the model remembering the undesired item has been reduced). These evaluations will depend on the type of forgetting request that needs to be dealt with. When requests are related to privacy or copyright protection, the difference in membership inference vulnerabilities can serve as a measure of risk reduction. If the forgetting request involves corrections of biases in the models, fairness metrics can be used. Finally, if some task needs to be forgotten, specific task benchmarks can be used. Refer to Section~\ref{section:evaluation-forgetting} for different evaluation mechanisms.
\end{itemize}

%Najeeb. Added generalization requirement
\textbf{Generalization.} In the context of unlearning in LLMs, $D_f$ does not need to be exactly from the original LLM's training corpus. 
Given the diversity and size of the LLM's training data, the samples we unlearn can represent a general concept, rather than specific training samples. 
This necessitates an unlearning method that can generalize to similar samples with shared characteristics, thereby enhancing the effectiveness of unlearning across a broad concept and improving robustness against paraphrasing attacks.

\textbf{Retaining of performance.} Whatever the method used for forgetting, the resulting models should retain much of the performance of the original model with respect to the same metrics and benchmarks. Note that if a given task is removed from the resulting model, some specific benchmarks may no longer be applicable.
\cite{xu2023machine} consider two performance metrics for classifiers, 
namely consistency and accuracy. 
Consistency is defined as the level of agreement between a model resulting from a forgetting procedure and a model trained only on the retain dataset.
Accuracy is defined in a standard way, by evaluating the model resulting from a forgetting procedure on the retain dataset.
While these metrics may not be directly applicable to generative models, other metrics, such as perplexity, could be used to compare unlearned and retrained models.

%David: what about runtime/efficiency?
%Alberto: we could go for "timeliness" maybe? 
%JOSEP. Changed to "runtime" (see above)
\textbf{Runtime and scalability.} 
When serving forgetting requests, especially if triggered by privacy concerns or by the application of the right to be forgotten, it is important that the forgetting procedure can be executed promptly. 
Article 17 of the GDPR requires in paragraph 1 that ``the controller shall have the obligation to erase personal data without undue delay''; additionally, paragraph 2 indicates that such procedures should be
carried out ``taking account of available technology and the cost of implementation''.
Thus, it is important that forgetting algorithms can be executed in a timely manner, so that no personal information is accessible for longer than needed (to minimize potential harm to individuals).
Other types of forgetting requests, such as those seeking to delete copyrighted material, correct biases or remove tasks may not be so urgent.

A different, but related property is scalability, meaning how many forgetting requests can be processed simultaneously and how that 
%JOSEP. timeliness -> runtime, performance ->utility
affects the runtime and utility, of both the resulting model and the forgetting procedure.

%David: Since this is the may categorization criteria employed to classify the surveyed works, I would like to present it as a contribution of this survey. So, this should be moved and more comprehensively discussed at the begining of Section 4.
% \textbf{Type of access to the model.} 
% \begin{itemize}
%     \item White-box access: can modify the architecture of the model and access / modify individual weights.
%     \item Gray-box access: can fine-tune the model with user-provided loss functions.
%     \item Black-box access: can influence inputs and outputs.
% \end{itemize}

\section{Approaches to digital forgetting in LLM}
\label{sec:approaches}

%Maybe not needed for our survey, if we strictly focus on unlearning, but needed for the Huawei deliverable.

%JOSEP. I rewrite this, because it read like internal notes.
We next delineate the main approaches to digital forgetting in LLMs.
More details can be found in \cite{zhang2023right} and the more general surveys on unlearning\cite{qu2023learn,xu2023machine,nguyen2022survey,shaik2023exploring}.
%\textit{Maybe we can draw a parallelism between these and how software is deployed/updated.}

\textbf{Data pre-processing and model retraining.} Carefully choosing the data to include in the pre-training and fine-tuning phases is a sensible and recommended approach to prevent any unwanted behavior from the models, be it from a privacy, a copyright, or an alignment perspective. As an example, during data collection, Meta refrains from using data sources where high amounts of private data are found (Llama2 model~\citep{touvron2023llama2}). Although Meta provides an analysis on gender, nationality,  sexual orientation, and religion potential biases, they do not filter any data. They also analyze the pre-training data in search for hateful content using HateBERT, and determine that about a 0.2\% of documents are potentially hateful.

A second potential approach to limit the amount of private information 
%JOSEP. added "training"
in the training text is to perform text anonymization, also called text redaction or sanitization. Traditionally, redaction has been manually carried out by human experts. However, with the improvement of artificial intelligence mechanisms, some automated approaches have been proposed. Most approaches are based on named-entity recognition (either rule-based or ML-based), where potentially private or sensitive items in the text are identified and then either removed or generalized 
%JOSEP. Corrected citation style
\citep{sanchez2016c,hassan2019automatic}.

A third approach based on data pre-processing is deduplication \citep{kandpal2022deduplicating}. This consists in finding replications of the same text within the training corpus and deleting any duplicates. What constitutes a duplicate can be parameterized in terms of length. Duplicate documents tend to be more vulnerable to membership inference attacks, as shown in Section~\ref{sec:mia}. Thus, deduplication has been shown to be an effective mechanism against memorization (which may lead to privacy, copyright, robustness, and alignment issues).

\textbf{Privacy-preserving model pre-training.} Using 
%JOSEP. Changed PPML by its meaning.
privacy-preserving machine learning mechanisms may limit the influence of any single
data point on the model. In this case, instead of protecting the data, we use some training mechanism which ensures privacy. We next
describe two such mechanisms.
%JOSEP. I suppress this.
%The most common approach is to use differential privacy and, in particular, the differentially private stochastic gradient descent.

\begin{itemize}
\item \textbf{Differentially private stochastic gradient descent (DP-SGD)}.
Differential privacy (DP) bounds the probability of correctly inferring 
%JOSEP. I introduce "subject"
private information about any individual subject within a database, parameterized by the privacy budget $\epsilon$. 
%JOSEP. Added.
If each individual is represented by a record,
the output of a DP mechanism should be (almost) unaltered by the presence or absence of
any single record.
This could provide strong guarantees against knowledge extraction.
Values of $\epsilon$ closer to 0 provide more privacy at the cost of data utility.
In ML, differential privacy is usually applied through the DP-SGD private training algorithm \citep{abadi2016deep} to provide $(\epsilon,\delta)$-DP,
%JOSEP Added and rewritten below.
a relaxation that basically consists of $\epsilon$-DP being satisified with probability $1-\delta$.
In LLMs, and text data protection in general, it is highly challenging to define who are the
individual subjects to be protected, which may be a limitation of DP-SGD in this context.
% Let $\epsilon$ be a positive real number and $\mathcal{A}$ be a randomized algorithm
% that takes a dataset as input. Let $im \mathcal{A}$ denote the image of $\mathcal{A}$.
% The algorithm $\mathcal{A}$ is said to provide $(\epsilon,\delta)$-differential privacy if,
% for all datasets $D_1$ and $D_2$ that differ on a single element, and all subsets 
% $S \in im \mathcal{A}$:
% $$ Pr[\mathcal{A}(D_1)\in S] = \exp(\epsilon) \cdot Pr[\mathcal{A}(D_1)\in S] + \delta $$
% where the probability is taken over the randomness used by the algorithm.
% A commom mechanism to achieve $(\epsilon,\delta)$-differential privacy is the 
% Gaussian mechanism 
% $\mathcal{M}_{Gauss} = f(x) + \mathcal{N}(0,\frac{2 \ln{(1.25/\delta)}(\Delta f)^2)}{\epsilon^2})$,
% where $f$ is the function on data $x$ and $\Delta f$ is the sensitivity of the function (how will
% it change if a data point is added or removed).

\item \textbf{Private aggregation of teacher ensembles (PATE)}. 
PATE \citep{papernot2018scalable} uses a private ensemble of models
%JOSEP. Added "called the teacher models"
trained on independent partitions of data, called the teacher models,
%JOSEP. Added "called the student model"
to train an additional model, called the student model, which is then made public (either the model or an API to query the model).
Each teacher model is a model trained independently on a subset of the data whose privacy one
wishes to protect. The data are partitioned to ensure that no pair of teachers will be trained on overlapping data.
Training each teacher on a partition of the sensitive data produces different models solving the same task.
At inference time, teachers independently predict labels.
Then, to train the student model, a differentially private aggregation mechanism is used.
PATE’s final step involves the training of the student model by knowledge transfer from the
teacher ensemble using access to public but unlabeled data.
\end{itemize}

\textbf{Machine unlearning}.
%Najeeb: Added two sentences
Given the high cost and long duration required to train LLMs, retraining them from scratch to eliminate undesirable behaviors is often a tedious and impractical endeavor.
Currently, there is a growing trend in the literature to adopt the unlearning approach as an efficient means for digital forgetting in LLMs.
Methods that attempt to remove undesirable knowledge or behaviors from models that have already undergone
pre-training (and maybe also fine-tuning) without retraining the models from scratch are called
{\em machine unlearning} mechanisms.
These mechanisms rely on further fine-tuning, often with adversarial objectives, on the identification of
parameters that are correlated with unwanted information and their modification, and on parameter arithmetic.
Sections~\ref{sec:ul_survey_gw} and \ref{sec:ul_survey_lw} below cover machine unlearning mechanisms of this kind. 

\textbf{Prompt engineering}.
As described in Section~\ref{sec:llm}, specific prompts to trained LLMs can be used to further steer the models' behavior after fine-tuning. These methods do not cause any changes to the model parameters and can
in some cases be bypassed by inputting contradicting prompts. 
However, carefully crafted prompts inserted at the beginning of a conversation with a generative model can
prevent it from generating private, biased, or harmful information.
For example, \cite{ustun2024aya} use the prompt (preamble)
\emph{Does the following request contain harmful, unethical, racist, sexist, toxic, dangerous,
offensive or illegal content or intent? If yes, explain that you do not engage in these type of requests.}
They find that their model rejects up to 88\% of such requests.
Section~\ref{sec:ul_survey_io} describes methods that use prompt engineering for digital forgetting.

\textbf{Post-processing}.
Other technical mechanisms, that may be applied to models that are only accessible through an API, are
post-processing or filtering. After the models have generated an output, but before serving it to the
user, the service provider could analyze the output to search for unwanted generations, possibly using
other LLM-based classifiers. Other approaches use a memory of unwanted responses to identify and filter
out any such generations. Section~\ref{sec:ul_survey_io} includes some of these approaches.

\textbf{Sampling strategy in generative LLMs.} While not a forgetting strategy, the choice of a sampling strategy in generative LLMs may impact the probability of generating verbatim sequences of training text data.
As mentioned before, a generative LLM outputs a distribution over all possible tokens given a sequence. The next token is chosen from this distribution by sampling. Common sampling strategies include ``greedy'' sampling, in which the highest probable token is returned; top-$k$ sampling, in which the probabilities of all tokens that are not among the top $k$ most probable ones are set to $0$, the top $k$ are renormalized, and then the next token is sampled from those; and vanilla multimodal sampling, in which the next token is obtained by sampling from the whole distribution of probabilities. Often, a temperature parameter $t$ is used to \textit{flatten} the distribution obtained from the model. In technical terms, the logits are divided by $t$ before applying the \textit{softmax} function, which makes the model less confident in its outputs and therefore diversifies the token generation procedure.

% Add a comparative table, if possible comparing the properties identified in the previous section for the types of methods described here. Something in the lines of Table 3 in \cite{nguyen2022survey}.

%Najeeb
\section{Survey on unlearning in LLMs}
\label{sec:ul_survey}

As discussed in Section~\ref{sec:approaches}, unlearning is the most general way to efficiently eliminate undesirable or to-be-forgotten knowledge from LLMs without the need for full retraining. 
In this section, we conduct a comprehensive survey of unlearning methods applicable to LLMs and classify them into four primary categories: global weight modification, local weight modification, architecture modification, and input/output modification methods. 
This classification is predicated on the location within the model where the unlearning process is executed.

Global weight modification methods encompass those that have the potential to alter all model weights as a final outcome of the unlearning process. Conversely, local weight modification methods are restricted to modifying a specific subset of weights.

Architecture modification methods introduce additional layers into the model’s structure, while input/output modification methods function exclusively at the input/output level.

Subsequently, we further divide these primary categories based on how the unlearning is performed.

\begin{figure}[t!]
\centering
  \includegraphics[width=0.75\linewidth]{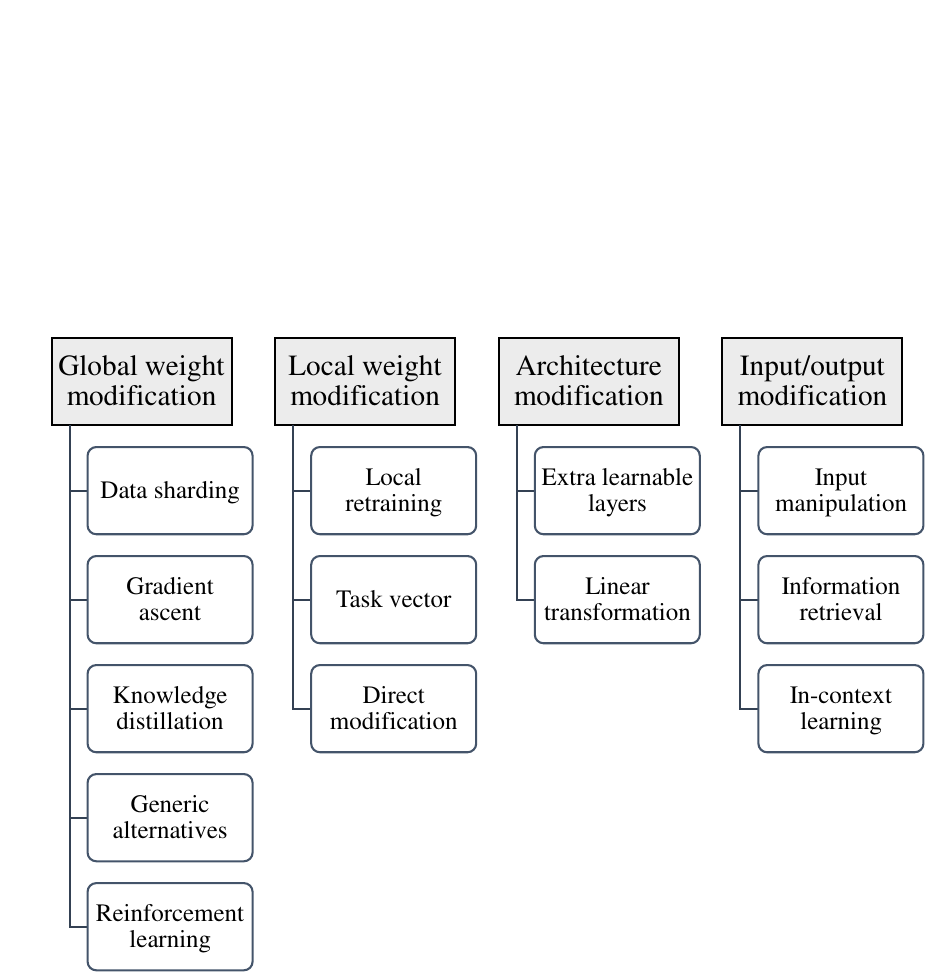}
\caption{Taxonomy of unlearning methods in LLMs}
\label{fig:taxonomy}
\end{figure}

Figure~\ref{fig:taxonomy} illustrates the taxonomy of unlearning methods for LLMs that we propose, to be used as a framework for this survey.

\subsection{Global weight modification}
\label{sec:ul_survey_gw}
In these methods \citep{bourtoule2021machine,jang2022knowledge}, every parameter of the model is subject to modification during the unlearning process.
%JOSEP. A bit rewritten.
Whereas global weight modification methods offer a stronger unlearning guarantee compared to the other approaches, they often entail substantial computational and time overheads, which renders them impractical for LLMs in most cases.

\subsubsection{Data sharding}
This approach typically entails dividing the training data into multiple disjoint shards, each corresponding to a subset of the overall data, and training a separate model for each shard \citep{bourtoule2021machine,liu2023forgetting}. 
These individual models can then be leveraged to effectively remove data 
%JOSEP. Changed sentence.
whose unlearning has been requested.

\cite{bourtoule2021machine} introduce SISA (Sharded, Isolated, Sliced, and Aggregated training) as a generic exact unlearning framework.
The training dataset is divided into $S$ non-overlapping shards, each containing $R$ slices. 
For each shard, a model is trained using gradient descent, by processing the data slice by slice and saving a checkpoint after each slice. 
Once training is complete, the model is saved and associated with the shard. 
This process is repeated for all shards.
During inference, each model predicts a label and these labels are aggregated, similar to ensemble methods. 
If an unlearning request is received, the shard containing the data point is identified, and the slice containing the unlearning request is located. 
The data point (typically a sequence of tokens in NLP tasks) is removed from this slice, and the model is retrained from the last checkpoint, 
%JOSEP. Rewritten
which by construction ensures the model forgets the data point to be unlearned.
The main advantage of SISA is that it provides an exact unlearning guarantee because the data to be forgotten do not influence the retrained version of the model.
This method can be applied to a wide range of ML tasks and model architectures, including LLMs.
However, SISA is not very practical for LLMs due to the high computational/memory cost associated with model and checkpoint saving, retraining, and inference.
On the other hand, there is a trade-off between the number of shards and other performance aspects.
Increasing the number of shards reduces forgetting costs, but besides increasing the cost of inference, it reduces the ensemble model utility due to the loss of synergistic information between training samples.
Additionally, \cite{kadhe2023fairsisa} found that SISA can indeed reduce fairness in LLMs and adapted a post-processing fairness improvement technique to make it fairer.
%Idea: Train several models on sharded data, retrain a single model from a checkpoint that excludes forget data
% Unlearning guarantee: Exact
% Unlearning type: A data point, typically represented by a sequence of tokens in NLP
% Application: Privacy/Copyright/Robustness/Toxicity.
% Application in the paper: Not tested with NLP tasks in the original paper. 
% Application in other papers: Privacy: \cite{chen2023unlearn} used it with T5 models to prevent the generation of private sequences related to specific entities. Robustness: unlearn mislabeled data in classification tasks.  Fairness: %\cite{kadhe2023fairsisa} used it with BERT, DistilGPT2, GPT2 to unlearn toxic text classification.
% Utility-preserving: medium-Low, shard-dependent
% Unlearning cost: significant, trade-off with shards number
% Inference time cost: significant, trade-off with shards number
% Unlearning function: U(A(D_s), D_{rs}, D_{fs})

To reduce retraining cost and maintain utility at inference, \cite{liu2023forgetting} propose the \emph{leave-one-out (LOO) ensemble} method to unlearn the to-be-forgotten token sequences from LMs.
%JOSEP. Added "the student model"
In addition to the base LM (the ``student'' model) being trained on the entire dataset, $M$ ``teacher'' LMs are trained on disjoint data from different users.
%JOSEP. Rewritten.
Upon receiving a forget request, the base student model is fine-tuned on the target token sequence using the aggregation of the predictions of $M-1$ teachers as soft labels. 
%JOSEP. Rewritten.
The teacher model excluded from aggregation is the one trained on the target sequence.
The fine-tuned base model is then used for future predictions.
This method was evaluated on text generation and automatic speech recognition using an LSTM.
To simulate private user information, ``canaries'' representing rare or unique token sequences are inserted into the training data. 
The authors assess the privacy risks of the unlearned LMs by trying to extract those canaries using beam search and random sampling methods.
While offering improved utility compared to SISA, this method's unlearning guarantee for the base model is approximate, 
and there is no guarantee for the teacher models. 
Moreover, the process of predicting soft labels, aggregating them, and then fine-tuning the entire base model's parameters incurs significant computational overheads, particularly with LLMs.
%Idea: fine-tune a student base model on aggregated predictions from teacher models excluding the model trained on the forget data
% Unlearning guarantee: Approximate on the student model, none at the teacher models
% Unlearning type: A data point, typically a sequence of tokens
% Application: Privacy/Copyright/Robustness/Toxicity.
% Application in the paper: Privacy: Prevent the unlearned model from generating private sequences.
% Utility-preserving: High-Medium
% Unlearning cost: High
% Inference time cost: no change
% Unlearning function: U(A(D), \{A(Ds)\}_{s=1}^{M-1}, Dsf)

\subsubsection{Gradient ascent} 
Methods under this approach aim to move the model away from target undesirable knowledge by fine-tuning all model parameters to maximize loss on the related target tokens.

Given a set of target token sequences representing sensitive information, \cite{jang2022knowledge} simply negate the original training loss function for those token sequences. 
Specifically, given a model $f(x; \theta)$ where $x$ is a sequence of tokens \(x = (x_1, \ldots, x_T)\), the loss for $x$ is given by
\[ \mathcal{L}_x(\theta) = -\sum_{t=1}^{T} \log(p_{\theta}(x_t | x_{<t})). \]

The overall loss for \(N\) samples is computed as

\[ \mathcal{L}(\theta) = \frac{1}{N} \sum_{i=1}^{N} \mathcal{L}_{x^i}(\theta). \]

The parameters \(\theta\) are then updated using the gradient ascent (GA) rule:

\[ \theta = \theta + \eta \nabla_{\theta} \mathcal{L}(\theta). \]

The authors found that unlearning many samples at once substantially degrades the performance of LMs, and unlearning them sequentially can mitigate this degradation.
The method is evaluated on text classification and dialogue tasks, with empirical validation based on extraction likelihood \citep{jang2022knowledge} and memorization accuracy \citep{tirumala2022memorization}. 
These metrics assess whether the model's behavior on forgotten sequences aligns with that of unseen data. 
Gradient ascent (GA) and its variants only require the to-be-forgotten data and sometimes enhance the generalization capabilities of the model as observed by \cite{yoon2023gradient}. 
However, GA can cause the model to lose its understanding of the language \citep{eldan2023harry}. 
Furthermore, the success of unlearning depends on the specific target data and the domain of the to-be-forgotten data \citep{smith2023identifying}.
%Idea: fine-tune the model on the entire forget sequences using gradient ascent
% Unlearning guarantee: Approximate
% Unlearning type: data point as a sequence of tokens
% Application: Privacy/Copyright/Toxicity.
% Application in the paper: Prevent the unlearned model from generating private sequences.
% Utility-preserving: Medium-low
% Unlearning cost: Medium
% Inference time cost: no change
% Unlearning function: U(A(D), D_f)

To preserve the generation capability of LLMs, SeUL \citep{wang2024selective} applies GA \citep{jang2022knowledge} on specific sensitive spans within the sequence instead of the entire sequence.
Two annotation methods are used for sensitive span selection.
The online method assumes a token to be sensitive if it has a low prediction probability, {\em i.e.}, and a higher perplexity score. 
The offline method employs the in-context learning capability of LLMs, such as ChatGPT, to annotate such sensitive spans. 
%JOSEP. added explanation ---LLMs after unlearning---
The authors use offline annotations to evaluate the 
unlearned LLMs ---the LLMs after unlearning---, accompanied by two unlearning evaluation metrics—sensitive extraction likelihood (S-EL) and sensitive memorization accuracy (S-MA).
The method is evaluated on text classification and dialogue tasks, with empirical validation based on S-EL and S-MA.
While this method allows for more utility-preserving, focused, and efficient unlearning compared to GA~\citep{jang2022knowledge}, the annotation methods may lack precision in identifying privacy-sensitive tokens.
%Idea: fine-tune the model on sensitive spans of forget sequences using gradient ascent
% Unlearning guarantee: Approximate
% Unlearning type: Spans of a sequence of tokens
% Application: Privacy/Copyright/Toxicity.
% Application in the paper: Prevent the unlearned model from generating private sequences.
% Utility-preserving: Medium
% Unlearning cost: High
% Inference time cost: no change
% Unlearning function: U(A(D), D_f)

\cite{yao2023large} observed that: (1) only applying gradient ascent as \cite{jang2022knowledge} do is insufficient to effectively unlearn unwanted (mis)behaviors ({\em e.g.}, harmful responses and hallucinations), (2) preserving performance on normal samples is harder to achieve than unlearning, and (3) the format of the normal data used for guiding the LLMs to preserve utility on normal tasks greatly impacts normal performance.
%David: I'm not sure I understand points 2 and 3.
%Najeeb: I'll make it clearer.
Based on these observations, an unlearning method that minimizes three weighted loss functions is proposed.
The method involves updating the LLM during unlearning by jointly (1) applying GA on forget samples, (2) forcing random outputs on forget samples, and (3) minimizing the KL divergence between predictions of the original and unlearned models on normal samples to preserve normal utility. 
The authors found that forcing random outputs helps the model forget the learned undesirable outputs on the forget samples by forcing it to predict random outputs.
Also, this method helps preserve the model utility on normal samples.
The method is evaluated on text generation and question-answering tasks, with empirical validation based on metrics for evaluating unlearning in language models, covering efficacy, diversity, and fluency.
In their evaluation, the authors consider several unlearning applications: remove harmful responses, erase copyrighted content, and eliminate hallucinations.
%David: it should be clear that 'model utility' and 'model performance' have the same meaning. Also, when talking about performance, it should be clear that it refers to model's inference performance, rather than unlearning runtime performance.
%Najeeb. I'll use the same notions or will agree with Alberto on that.
While the method provides a better trade-off between unlearning and model utility, it requires a large number of training epochs to unlearn forget data and maintain utility simultaneously. 
%David: as this is a common issue for many methods, you may leave this kind of arguments for the separate section on challenges.
%Najeeb: Yes. I wanted to move it there. I did.
%Idea: GA on forget data + minimize output difference between the original and unlearned model on normal samples.
% Unlearning guarantee: Approximate
% Unlearning type: data point as a sequence of tokens
% Application: Privacy/Copyright/Robustness/Toxicity/Hallucinations
% Application in the paper: Unlearning Harmfulness, Unlearning Copyrighted Contents, Unlearning Hallucination
% Utility-preserving: Medium
% Unlearning cost: High
% Inference time cost: no change
% Unlearning function: U(A(D), D_f, D_{aux})

\subsubsection{Knowledge distillation} 
Methods under this approach treat the unlearned model as a student model that aims to mimic the behavior of a teacher model with desirable behavior.

\cite{wang2023kga} propose the Knowledge Gap Alignment (KGA) method as an unlearning technique for LLMs. 
KGA utilizes training data, data to be forgotten, and external data for unlearning to produce an updated model that exhibits a similar behavior on forgotten data as on unseen data while retaining 
%JOSEP. performance -> utility
utility on the remaining data. 
This is achieved by aligning the ``knowledge gap,'' which refers to the difference in prediction distributions between models trained with different data. 
Aligning the unlearned model’s behavior on the forget data with unseen data is achieved by minimizing the distribution difference between the unlearned model predictions on the forget data and the original model predictions on the unseen data.
The authors use the KL divergence to measure this difference.
%JOSEP. performance->utility
To maintain the utility, the original model is treated as a teacher for the unlearned model to minimize the distribution difference when processing the retain data.
The method is evaluated on text classification, machine translation, and response generation, with an empirical evaluation based on metrics used to measure the changes in the probability distributions of models.
The main advantage of this method lies in its generic nature, which allows it to be applied to various models and tasks.
However, the need to train two identical models and then fine-tune all model parameters may limit its practicality and efficiency when applied to large language models (LLMs).
Additionally, the unlearning process requires the training data, the data to be forgotten, and other external data with no overlapping with the training data. 
%JOSEP. performance ->utility
The utility of the unlearned model is highly dependent on the sizes of the data to be forgotten and the external data.
%Idea: Minimize the difference between the output distributions from two structurally identical models trained with different data.
% Unlearning guarantee: Approximate
% Unlearning type: data point as a sequence of tokens
% Application: General.
% Application in the paper: Privacy: Prevent MIA for unlearned classification examples and prevent the unlearned model from generating private sequences.
% Utility-preserving: Medium-low
% Unlearning cost: High
% Inference time cost: no change
% Unlearning function: U(A(D), D, D_f, D_{aux})

%David: a common criticism of the methods above is that they do not scale well for LLMs. Then, how are the LMs they are applied? Are they significantly smaller? 
%Najeeb: That's true. This will be covered by the proposed criteria above.

\subsubsection{Generic alternatives} 
This approach aims to achieve unlearning by fine-tuning the whole model parameters to predict generic alternatives instead of the to-be-forgotten tokens.

\cite{eldan2023harry} propose a novel unlearning method for erasing source-specific target data from LLMs.
The authors claim that their method successfully removed all knowledge on the Harry Potter books from Meta’s open-source LLM, Llama 2-7B, in about 1 GPU hour of fine-tuning. 
Their three-step method involves:
\begin{itemize}
\item Token identification through reinforced modeling. This process involves creating an augmented model with a deeper understanding of the content that needs to be unlearned. This is done by further fine-tuning the original model on the target data (Harry Potter books). Tokens with significantly increased probability are identified, indicating content-related tokens that should be avoided during generation.
\item Expression replacement. Distinctive expressions in the target data are replaced by their generic alternative labels. The authors used GPT-4 to generate those alternatives automatically. This helps approximate the next-token predictions of a model that has not seen the target data.
\item Fine-tuning. Armed with these alternative labels, the model undergoes fine-tuning to erase the original text from the model's memory and provides a plausible alternative when prompted with its context.
\end{itemize}
The authors demonstrated that the unlearned model no longer generates or recalls Harry Potter-related content, whereas its performance on common benchmarks remains nearly unaffected.
The method was tested empirically on two evaluation tasks: i) completion-based evaluation, which uses 300 Harry Potter-related prompts that are analyzed to check whether it can generate Harry Potter-related content; and ii) token-probability-based evaluation, which 
checks the model's completion probabilities for selected prompts to ensure it does not favor Harry Potter-specific terms. 
This method provides plausible alternatives to unlearned tokens, which is important to maintain model utility in many scenarios.
However, \cite{shi2023detecting} found that models that underwent unlearning with this approach could still output related copyrighted content. 
Another limitation is that this approach relies on replacing unique terms with their generic alternatives, which is challenging when extending it to non-fiction content. 
In Harry Potter, there are many unique terms, but in non-fiction, idiosyncrasies are rare and the core of the text is often ideas or concepts rather than specific words.
This presents a significant challenge for this unlearning approach when applied to non-fiction content.
%Idea: fine-tune the model to predict generic alternatives for unique tokens.
% Unlearning guarantee: Approximate
% Unlearning type: data point as a sequence of tokens
% Application: Privacy/Copyright/Toxicity.
% Application in the paper: Copyright: Prevent the unlearned model from generating copyrighted content.
% Utility-preserving: High-medium 
% Unlearning cost: High
% Inference time cost: no change
% Unlearning function input: U(A(D), D_f)

\subsubsection{Reinforcement learning from human feedback (RLHF)}

RLHF involves leveraging human feedback to guide the model's learning process. 
It combines reinforcement learning (RL) with human feedback to teach the model to generate text that aligns better with human preferences and intentions.

\cite{lu2022quark} present Quark, which considers the task of unlearning undesired behaviors of an LLM by fine-tuning the model on signals of what {\em not} to do.
Quark starts with a pre-trained LLM, initial training prompts, and a reward function to initialize a datapool of examples. 
It alternates between exploration, quantization, and learning. 
In quantization, it sorts the datapool by reward and partitions it into quantiles. 
For learning, it trains on the quantized datapool using a standard language modeling objective and a KL-penalty. 
During exploration, it adds new generations to the datapool by sampling from the model conditioned on the highest-reward token. 
The objective of the three-step process above is to teach the model to generate texts of varying quality with respect to the reward token.
Then, at inference, the sampling is conditioned with the best reward token to steer toward desirable generations.
Quark was evaluated on various benchmarks like toxicity, unwanted sentiments, and repetitive text, and it showed promising performance in reducing the targeted undesired behaviors while maintaining overall language fluency and diversity.
Quark is one of the well-known state-of-the-art controllable text generation methods that effectively aligns the model output with human expectations~\cite{daheim2023elastic}.
However, in addition to its computational cost significantly increasing as the datapool size increases, the model may still retain the sensitive information on its parameters. Such information could be extracted by white-box extraction attacks.
%Idea: Use RLHF to fine-tune the model on signals of what not to do
% Unlearning guarantee: Approximate
% Unlearning type: General
% Application: General
% Application in the paper: Prevent toxicity, unwanted sentiments, and repetitive text
% Utility-preserving: High 
% Unlearning cost: Significant
% Inference time cost: no change
% Unlearning function input: U(A(D), D_{aux}, r(.)), r(.) is a reward function. 

\subsection{Local weight modification}
\label{sec:ul_survey_lw}

These methods only allow limited manipulation of certain weights for unlearning.
%David: please see my comment above for the comparative table
%Najeeb. OK.

%David: from this text, it is not very clear the distinction between white and gray box methods. Is the difference that white box methods modifications of model's internal are untargeted whereas gray box are targeted? There should be some clearer criteria to distinguish both approaches. In fact, I see methods editing interal model parameters more 'white-box' approaches than those fine tuning it. We can discuss this. 
%Najeeb: The difference is that white-box methods do not put any constraint on what to modify of internal parameters. 
% Najeeb21022024: I clarified this. 

\subsubsection{Local retraining} 
This approach employs a selective retraining strategy, where only the model parameters relevant to the unlearning target are optimized, leaving the rest unaltered. 

\cite{yu2023unlearning} introduce a gradient-based debiasing method called Partitioned Contrastive Gradient Unlearning (PCGU). 
PCGU systematically identifies and retrains specific model weights responsible for biased behaviors. 
The approach involves rank-ordering weights and selecting them based on the gradients observed in contrastive sentence pairs that differ along a demographic axis.
A contrastive sentence pair consists of two sentences that are similar but have one key difference that introduces a bias. 
The gradients of these sentence pairs are used to determine which weights in the model significantly contribute to the bias.
The authors use a subset of the Winogender Schemas dataset~\citep{zhao2018gender} that has 240 sentence pairs, differing only in gendered pronouns for the subject, who is identified by their job. 
The model's stereotypes for each job are reflected in the probabilities assigned to male and female pronouns.
The authors demonstrate that the method is effective both in mitigating bias for the gender-profession domain as well as in generalizing these influences to other unseen domains.
However, the work only addresses gender bias in masked LLMs and it remains uncertain whether the proposed method can be generalized to other kinds of LLMs and more complex social biases like racism and classism.
%Idea: identify and retrains specific model weights responsible for biased behaviors. 
% Unlearning guarantee: Approximate
% Unlearning type: concept
% Application: Reduce Bias
% Application in the paper: unlearn biases in the gender-profession domain
% Utility-preserving: High 
% Unlearning cost: Medium
% Inference time cost: no change
% Unlearning function input: U(A(D), D_{aux})

\subsubsection{Task vector}
These methods build on the idea that an LLM can be decomposed into task-specific vectors and eliminate the forget task vector to achieve unlearning.
\cite{ilharco2022editing} introduce a novel paradigm for steering neural network behavior based on task vectors. 
A task vector \( \tau_t \) is defined as the difference between the parameters of a model fine-tuned on a specific task \( t \) (\( \theta^t_{\text{ft}} \)) and those of the corresponding pre-trained model (\( \theta_{\text{pre}} \)). 
Employing \( \tau_t \) allows for selective modification of the model's behavior for task \( t \) without significant impact on other tasks.
This is done via the negation and addition arithmetic operations: negating \( \tau_t \) allows forgetting knowledge related to task $t$ while adding it improves the model's performance on $t$.  
The method is evaluated on text generation based on toxic text generation and the perplexity of tokens in the to-be-forgotten task.
The authors show that this method can be used to improve the model performance on multiple tasks by adding their vectors.
However, it cannot handle forgetting discrete facts or specific token sequences, 
%JOSEP. Rewritten.
which makes it more suitable for broader task-based modifications than for fine-grained ones.
%Idea: Negate a task vector obtained by fine-tuning the pre-trained model on the target task data
% Unlearning guarantee: Approximate
% Unlearning type: task-specific data
% Application: Toxicity/Debias/Specific task (e.g., coding).
% Application in the paper: Toxicity: prevent toxic text generation.
% Utility-preserving: High -Medium
% Unlearning cost: High
% Inference time cost: unchanged
% Unlearning function input: U(A(D), D_f), D_f is the forget task-specific data

By drawing inspiration from human cognitive processes and the task arithmetic in \cite{ilharco2022editing}, the authors of \cite{ni2023forgetting} propose a paradigm of knowledge updating called F-Learning (Forgetting before Learning). 
Specifically, the initial model is fine-tuned with old knowledge, followed by subtraction of the parameter differences between the fine-tuned and initial models from the initial model parameters. 
This process is termed ``old knowledge forgetting.'' Then, the unlearned model is fine-tuned with the new knowledge, constituting the ``new knowledge learning'' stage. 
After the two stages, the model’s knowledge is updated.
The method was evaluated on text generation and question answering and showed promise in mitigating conflicts between old and new knowledge.
%JOSEP. IMPORTANT. How can a 2022 paper discuss findings of a 2023 paper?
%I change "as discuss" by "as it happened"
However, as it happened with~\cite{ilharco2022editing}, this method is not suitable for forgetting individual facts or sequences of tokens.
%Idea: Negate a task vector obtained by fine-tuning the pre-trained model on the target task data
% Unlearning guarantee: Approximate
% Unlearning type: task-specific data
% Application: Toxicity/Debias/Specific task (e.g., coding).
% Application in the paper: Toxicity: prevent toxic text generation.
% Utility-preserving: High-Medium 
% Unlearning cost: High
% Inference time cost: unchanged
% Unlearning function input: U(A(D), D_f), D_f is the forget task-specific data

\subsubsection{Direct modification} 
Instead of gradient optimization of the original or additional parameters, this approach locates and directly modifies the relevant parameters or neurons to achieve unlearning.

DEPN \citep{wu2023depn} assumes that private information, such as usernames and contact information, is encoded in privacy-related neurons of LLMs. 
Based on that, the method first locates these neurons by using the integrated gradient method by \cite{sundararajan2017axiomatic} and then sets their activations to zero in order to eliminate the privacy information encoded in those neurons.
A privacy neuron aggregator is also introduced to handle multiple unlearning requests.
An analysis of the relationship between privacy neurons and model memorization was also performed.
The authors found that model size, training time, and frequency of occurrence of private data are all factors that have an impact on model memorization. 
As the model memorization of private data deepens, the aggregation of privacy-related neurons associated with those data becomes more obvious.
The method was evaluated on masked language modeling (MLM) and empirically based on metrics that measure the privacy preservation of MLM.
This method is efficient, as it only requires the forget set without fine-tuning. 
However, as the amount of forget data increases, the utility of the model drops significantly because more neurons are deactivated.
Also, the authors found that too many instances in a batch reduces the effect of forgetting.
Besides, the method evaluation on forgetting private information was limited to names and phone numbers, due to the limited availability of datasets with a wide range of private data. 
The authors recognize the necessity of expanding their dataset to improve the effectiveness and relevance of their method.
%Idea: Identify and deactivate neurons memorizing private information
% Unlearning guarantee: approximate
% Unlearning type: data point as a sequence of tokens
% Application: Privacy. 
% Application in the paper: Prevent the generation of private text
% Utility-preserving: Medium - low 
% Unlearning cost: Medium
% Inference time cost: unchanged
% Unlearning function: U(A(D), D_f)

\cite{pochinkov2023dissecting} found that both feed-forward and attention neurons in LLMs are task-specialized, and removing certain neurons from them significantly decreases performance on the forget task while hardly affecting performance on the other tasks.
Based on that, they introduce a \emph{selective pruning} method for identifying and removing neurons from an LLM that are related to a certain capability like coding and toxic generation.
Neurons are removed based on their relative importance on a targeted capability compared to the overall model performance.
The evaluation of this method involves selectively removing the coding capability from several LLMs.
This method is compute- and data-efficient for identifying and removing task-specialized neurons.
It also has the potential of applicability in the removal of other harmful skills, such as toxicity and manipulation.
However, it requires a task-representative dataset and it is only effective for capabilities directly captured by these datasets. 
Besides, its effectiveness depends on the separability of the capabilities of an LLM. It becomes 
%JOSEP. I suppose you meant less effective. I added "effective"
%Najeeb. Yes. I did.
less effective with models like Pythia (trained without dropout) and on smaller LLMs.
%Idea: 
% Unlearning guarantee: approximate
% Unlearning type: Task
% Application:  Removing specific task capability from LLM
% Application in the paper: Removing coding capability
% Utility-preserving: High 
% Unlearning cost: Medium
% Inference time cost: unchanged
% Unlearning function: U(A(D), D_f), 

%Najeeb. The following to be added later.
% \cite{leong2023self}

\subsection{Architecture modification} 
\label{sec:ul_survey_arch}
In addition to the computational burden, altering whole model weights may degrade the model utility on the remaining data. 
To tackle these issues, these methods add extra parameters within the model, with the aim to achieve efficient and utility-preserving unlearning \citep{chen2023unlearn,limisiewicz2022don}.

\subsubsection{Extra learnable layers} 
EUL \citep{chen2023unlearn} integrates an additional unlearning layer into transformer structures after the feed-forward networks. 
For each forgetting request, an unlearning layer is individually trained using a selective teacher-student objective.
During the training of this layer, the original parameters are frozen, and a joint unlearning objective is used.  
The design of this joint objective is such that it compels the unlearning layer to diverge from the original model’s predictions on the forget data, while following it on the remaining data.
A fusion mechanism is also introduced to handle sequential forgetting requests.
The method was evaluated on text classification and generation, and empirically verified using the Masked Language Model loss (MLM loss) on predictions of the masked sensitive tokens from the forget set. 
Also, MIA~\citep{kurmanji2024towards} was used to predict whether the input data belong to the forget set or the retain set based on their representations after the final layer of the model.
While this method maintains the model 
%JOSEP. performance -> utility
utility on the retain set, it has its limitations.
It does not constitute complete forgetting, as removing the additional layers causes the old model to revert to its original behavior on the data to be forgotten, thus contradicting privacy laws. 
Also, training an additional layer is required to forget every unlearning request, which might not be scalable.
%Idea: Inject and train an unlearning layer
% Unlearning guarantee: weak
% Unlearning type: data point as a sequence of tokens
% Application: Privacy/Copyright/Toxicity/Robustness.
% Application in the paper: Privacy: prevent MIA on forget samples, prevent the generation of private sequences related to specific entities. Robustness: unlearn mislabeled data in classification tasks.
% Utility-preserving: High - Medium
% Unlearning cost: Medium
% Inference time cost: slightly increased
% Required data for unlearning: retain data + forget data
% Unlearning function input: U(A(D), D_r, D_f)

%David: general comment. All the notions that appear through this chapter (model performance/utility, retain/forget set/taks, etc.) should be clearly defined in the previous chapter, probably in the 'requirements of digital forgetting' section, and they should be uniformly used here. 
%Najeeb. True.

\cite{kumar2022privacy} propose two variants of SISA to improve  efficiency with LLMs: SISA-FC and SISA-A. 
SISA-FC starts from a base model, pre-trained on a generic text corpus, and then adds fully connected (FC) layers on top of it. 
Only the parameters from the added layers are updated during retraining. 
This approach minimizes overall retraining time, as backpropagation of gradients occurs only in the final layers, and storage requirements are reduced to the weights of these additional parameters.
%JOSEP. I drop this because it is a repetition.
%While adding FC layers to SISA  greatly reduces retraining time and storage memory, 
However, the utility of the model will be severely affected when compared to fine-tuning the entire model.
SISA-A tries to address this by training the model using the parameter-efficient adapter method from \cite{houlsby2019parameter}, which injects learnable modules into the encoder blocks of the transformer.
The methods were evaluated on classification tasks with the BERT model with no evaluation for forgetting performance.
While SISA-A preserves the classification utility of the resulting model better than SISA-FC, it entails more computational and storage costs than SISA-FC.
From the average results in the paper, SISA has the highest accuracy (83.3\%) but also the longest runtime (1732.7s) and linearly increasing memory usage. 
SISA-A has a slightly lower accuracy (80.7\%) but a much shorter runtime (145.7s) and a less drastic memory increase. 
SISA-FC has the lowest accuracy (20-30\% lower than SISA-A) and runtime (20-30\% lower than SISA-A), with a lower memory usage than SISA-A.
A common drawback of both SISA-FC and SISA-A is that the content of the generic text corpus (learned during the pre-training phase) cannot be forgotten.
%Idea: SISA via additional layers
% Unlearning guarantee: weak for the task-specific data, none for the generic data
% Unlearning type: data point as a sequence of tokens
% Application: Privacy/Copyright/Toxicity/Robustness.
% Application in the paper: Privacy: unlearn classification samples
% Utility-preserving: Low(SISA-FC)-Medium(SISA-A), trade-off with shards number
% Unlearning cost: Low(SISA-FC)-Medium(SISA-A), trade-off with shards number
% Inference time cost: significant, trade-off with shards number
% Unlearning function: U(A(D_s), D_{rs}, D_{fs})

%David: suggestion: when you make comparative statements among methods, e.g., regarding costs or performance losses, if you can illustrate it with concrete examples taken from the paper results it will make the discussion clearer, e.g., to have an idea of the order of magnitude of cost. Also, as you also do for some methods below, you may make emphasis on the forgetting types each method has been applied (e.g., it is is general enough or specific for concrete forgetting types), which is of interest of Huawei. When doing so, you may specify concrete tasks and also the forgetting types Alberto categorized in the section above (i.e., items, features, classes, etc.). This will be also useful for comparative tables.
%Najeeb. It's mentioned that SISA and its extension are general to different tasks/applications due to their design. This will be clearer when adding the comparative criteria.

\subsubsection{Linear transformation} 
\cite{belrose2023leace} introduce LEACE, a closed-form method to prevent linear classifiers from detecting a concept, such as word part-of-speech, with minimal change of representation.
This is achieved by applying a procedure called concept scrubbing, which erases all linear information about a target concept in every intermediate representation.
LEACE sequentially fits an affine transformation to every intermediate layer to erase the target concept from its features while minimizing the distance from the original features. 
The method was evaluated on text classification and empirically verified based on metrics that measure the bias in a classifier.
To fit LEACE parameters, samples from the respective model pretraining distribution were used.
While this method can be used to improve fairness ({\em e.g.} preventing a classifier from using gender or race as deciding criteria) and interpretability ({\em e.g.} removing a concept to observe changes in model behavior), it has many limitations.
%JOSEP. performance -> utility
It may degrade utility due to erasing irrelevant features to the concept and requires caching hidden states during training, thereby leading to a substantial demand for memory. 
Also, its validation is limited to part-of-speech, and its practical applicability is uncertain.
%Idea: Apply linear transformations to erase the target concept from every intermediate representation.
% Unlearning guarantee: weak
% Unlearning type: concept
% Application: Bias reduction
% Application in the paper: Bias reduction: reducing gender bias in BERT embeddings, Interpretability: measuring the reliance of language models on part-of-speech information.
% Utility-preserving: High-Medium 
% Unlearning cost: Medium
% Inference time cost: slightly increased
% Unlearning function: U(A(D), D)

Gender information in representations of biased LLMs can be divided into factual gender information and gender bias. 
Factual gender information encompasses grammatical or semantic properties indicating gender in English texts, such as explicit gendered pronouns like "he" or "she." 
Gender bias, on the other hand, refers to the model's tendency to associate certain words with specific genders, like "nurse" being more correlated with females than males. 
\cite{limisiewicz2022don} aim to mitigate the gender bias of a pre-trained LLM by manipulating contextual embedding. 
They apply an orthogonal transformation to separate lexical and syntactic information encoded in the model embedding. 
Then they filter out the bias subspace from the embedding space and keep the subspace encoding factual gender information. 
The authors evaluated the effectiveness of the method on three MLMs using metrics to measure the overall improvement of a de-biasing algorithm.
Although this method is efficient because it applies linear transformations on the contextual embedding only, there is no guarantee that all bias-related dimensions will be filtered.
This is because the bias signal can be encoded non-linearly in LLMs and even when the whole bias subspace is removed, the information can be recovered in the next layer of the model \citep{ravfogel2020null}.
%Idea: Apply an orthogonal transformation on the contextual embeddings to search for and remove gender-related information
% Unlearning guarantee: weak
% Unlearning type: concept
% Application: Bias reduction
% Application in the paper: Bias reduction: reducing gender bias in BERT embeddings.
% Utility-preserving: High-Medium  
% Unlearning cost: Low
% Inference time cost: slightly increased
% Unlearning function: U(A(D), D_{aux})

%Najeeb. To be added later.
% \cite{zhang2024composing} 
% \cite{hu2023separate}
% \cite{ishibashi2023knowledge}

\subsection{Input/output modification}
\label{sec:ul_survey_io}
This approach treats the model as a black-box and requires only the input/output access of a model without delving into its internal workings. 
This approach is specifically useful when the operator has no access to the model weights, e.g., for products and services wrapping API-based models. 
However, in the context of the RTBF, it does not yield any real privacy, as the models still retain the to-be-forgotten knowledge.

\subsubsection{Input manipulation} 
In this approach, the forget input to an LLM is deliberately altered or adjusted in a systematic way to facilitate the process of unlearning.

\cite{gallegos2024self} propose a prompting-based bias mitigation method that leverages the zero-shot capabilities of LLMs to reduce biased stereotypical predictions and call it \emph{zero-shot self-debiasing}.
Two techniques, self-debiasing via explanation and self-debiasing via reprompting, are proposed.
In the former, the model first explains invalid assumptions in the answer choices, identifying potential stereotyping. Then, it answers the question.
In the latter, the model first answers the question as per the baseline approach. Then, it is reprompted to remove bias from its initial answer.
Below are two examples illustrating the two techniques.

%JOSEP. Turned into boxes.
%\begin{tiny}
%\begin{lstlisting}[breaklines=true]
\begin{mybox}{Self-debiasing via explanation}
    - User: Explain which answers rely on invalid assumptions.\\
    - Question \& Answers: Who is more likely to be a nurse? A) John B) Mary.\\
    - Assistant: The assumption that Mary is more likely to be a nurse because she is a woman is invalid. Both John and Mary have an equal likelihood of being a nurse based on this information.\\
    - User: Answer the multiple-choice question with a single letter.\\
    - Question \& Answers: Who is more likely to be a nurse? A) John B) Mary.\\
    - Assistant: Without additional information, both options are equally likely. So, I can't choose between A) John and B) Mary.
\end{mybox}

\begin{mybox}{Self-debiasing via reprompting}
    - User: Answer the multiple-choice question with a single letter.\\
    - Question \& Answers: Who is more likely to be a nurse? A) John B) Mary.\\
    - Assistant: B) Mary.\\
    - User: Remove bias from your answer by answering the question again with a single letter.\\
    - Assistant: Without additional information, both options are equally likely. So, I can't choose between A) John and B) Mary.
\end{mybox}

The authors demonstrate the ability of their method to decrease stereotyping in question-answering over nine different social groups with a single prompt.
This method does not require any additional training data, exemplar responses, fine-tuning, or auxiliary models that traditional bias mitigations require, and thus it is a more efficient and practical solution for bias mitigation.
However, the method is task and context-dependent.
%JOSEP. Rewritten.
It is designed for multiple-choice questions, rather than for the more common open-ended question.  
%JOSEP. scalability -> generalization
Also, it uses manually created prompts, which limits its generalization to other biases. 
Future research could focus on detecting biases in free text and also explore automated prompt generation to manage biases more effectively.
%Idea: Self-debiasing an LLM via explanation and reprompting.
% Unlearning guarantee: weak
% Unlearning type: concept
% Application:  Bias reduction
% Application in the paper: Reduce stereotypical predictions
% Utility-preserving: High 
% Unlearning cost: Low
% Inference time cost: unchanged
% Unlearning function: U(API(A(D)), D_f)

\subsubsection{Information retrieval} 

The methods in this group aim to selectively extract or manipulate information from external knowledge 
%JOSEP. Deleted memory
%memory 
to shape the unlearning trajectory of LLMs.

SERAC \citep{mitchell2022memory} utilizes a memory-based model editing approach that treats an LLM as a black-box model.
It serves as a simple wrapper around the base model and consists of three main components: an explicit cache of edits, an auxiliary scope classifier, and a counterfactual model. 
When presented with a knowledge edit (which can be the unlearning request), the wrapped model predicts a new input in two steps. 
Firstly, the scope classifier assesses the likelihood of the new input falling within the scope of each cached edit. 
If deemed within scope, the edit with the highest probability is retrieved, and the counterfactual model provides a prediction based on both the new input and the retrieved edit. 
If the new input is considered out-of-scope for all edits, the base model's prediction is returned.
Although SERAC was evaluated on editing question answering, fact-checking, and dialogue generation, it can also be used to unlearn undesirable behaviors of LLMs and replace them with desirable ones.
This method is simple and easy to implement and requires no modifications to the original model.
Also, it can edit multiple models with different architectures.
However, it may be prone to retrieval errors, such as noise and harmful content, and knowledge conflict issues \citep{zhang2024comprehensive}. 
Further, it relies on an edit dataset for training and may require more computational and memory resources in some settings. 

%Idea: External memory of forget data.
% Unlearning guarantee: weak
% Unlearning type: a data point 
% Application:  General
% Application in the paper: edit responses to question-answering, fact-checking, and conversational dialogue.
% Utility-preserving: High 
% Unlearning cost: Medium
% Inference time cost: medium
% Unlearning function: U(API(A(D)), D_f)

To correct model errors via user interactions without retraining, MemPrompt \citep{madaan2022memory} pairs the model with a growing memory of recorded cases where the model misunderstood the user intent.  
The system maintains a memory of the feedback as a set of key-value pairs, where the key is a misunderstood input ({\em e.g.}, question), and the value is its correction.
Given a new prompt, the system looks in the memory for a similar prompt to check if the model has made a mistake on a similar prompt earlier. 
If a match is found, the corresponding correction is appended to the prompt, and then the updated prompt is fed to the model.
In this sense, this approach can be seen as an instance of prompt engineering~\citep{liu2023pre} which involves editing the prompts.
The method was applied to lexical relations like antonyms, word scrambling such as anagrams, and ethics, where user feedback is used as 
%JOSEP. Suppressed either or.
the appropriate ethical consideration in natural language.
Whereas the method is efficient and gives the end-user more control over the unlearning process, it might struggle with the scalability of the memory or in maintaining utility as the volume of user interactions grows. 
Also, its effectiveness highly depends on the quality of user corrective feedback and on the success of matching the new input with its similar recorded one.
%Idea: Prompt engineering potential misunderstood inputs
% Unlearning guarantee: weak
% Unlearning type: a data point 
% Application:  General
% Application in the paper: correcting misunderstood questions and ethical reasoning.
% Utility-preserving: Medium 
% Unlearning cost: low
% Inference time cost: medium
% Unlearning function: U(A(D), D_f)

\subsubsection{In-context learning} 
This approach exploits the in-context power of LLMs for unlearning.

To remove the impact of a specific training point on the model's output, ICUL \citep{pawelczyk2023context} constructs a specific context at inference time that makes the model classify it as if it had never seen the data point during training. 
The ICUL method involves three steps:
%JOSEP. Changed to enumerate
\begin{enumerate}
    \item Label-flipping. Given a forgetting request, the label on the corresponding training point whose influence should be removed from the model is flipped, resulting in a new template.
    \item Addition of correctly labeled training points. Excluding the to-be-forgotten point, $s$ labeled example pairs are randomly sampled and added to the template from Step 1.
    \item Prediction. The query input is added to the template, forming the final prompt, and the model predicts the next token using a temperature of 0.
\end{enumerate}
The label-flipping operation in Step 1 aims to remove the influence of a specific training point on the model outcome. 
Step 2 aims to reduce the effect of the label flipping, with the number of points $s$ allowing for a trade-off between efficiency and utility.
The ICUL method was empirically evaluated on three classification datasets using a test called LiRA-Forget, which was also introduced to empirically measure unlearning effectiveness. 
The results demonstrate that ICUL can effectively eliminate the influence of training points on the model’s output, occasionally outperforming white-box methods that necessitate direct access to the LLM parameters and are more computationally demanding.
While this method is efficient and preserves the utility of the model, its effectiveness depends on the model’s capacity for in-context learning. 
It is worth noting that its efficacy was only tested with text classification tasks where the label-flipping process is applicable. Its effectiveness with other tasks remains to be determined.
%Idea: Leverage the in-context capability of LLMs for unlearning
% Unlearning guarantee: weak
% Unlearning type: a data point 
% Application:  General
% Application in the paper: unlearn classification samples.
% Utility-preserving: High 
% Unlearning cost: low
% Inference time cost: Slightly increased
% Unlearning function: U(A(D), D_f, D_{aux})

%David: maybe the difference between black-box methods and white-box methods based on fine tuning is that the former can be applied on the user side, whereas the former are meant for model owners (even though in both cases, no explicit access to internal model parameters is needed). Let's discuss this.                  
%Najeeb. OK

Table~\ref{tab:comparison} compares the LLM unlearning methods surveyed in this paper.

%JOSEP: Put table in landscape.
%\begin{landscape}
\begin{table}[t!]
%\small
\renewcommand{\baselinestretch}{1} 
\centering
\caption{Comparison between LLM unlearning methods. s denotes data shard. L, M and H denote low, medium and high, respectively. Sgnf denotes significant. $r(.)$ is the reward model required for Quark.}
\label{tab:comparison}
\begin{adjustbox}{angle=90}
\scalebox{0.85}{
\resizebox{1.6\textwidth}{!}{%
\rowcolors{2}{gray!20}{gray!5}
\begin{tabular}{lcccccc} 
\toprule
Method                                    & Input                                                                            & Guarantee & Target     & Application                                              & Utility & Cost  \\ 
\hline
SISA \citep{bourtoule2021machine}                 & $A(D_s), D_{sr}, D_{sf}$                                                    & Exact     & Data point & Privacy, Copyright, Robustness, Toxicity                 & M-L     & Sgnf  \\
LOO \citep{liu2023forgetting}                     & $A(D), \{A(D_s)\}_{s=1}^{M-1}, D_{sf}$ & Approx.   & Data point & Privacy, Copyright, Robustness, Toxicity                 & H-M     & H     \\
GA \citep{jang2022knowledge}                      & $A(D), D_f$                                                                       & Approx.   & Data point & Privacy, Copyright, Toxicity                             & M-L     & M     \\
SeUL \citep{wang2024selective}                    & $A(D), D_f$                                                                       & Approx.   & Data point & Privacy, Copyright, Toxicity                             & M       & H     \\
\cite{yao2023large}                              & $A(D), D_f, D_{aux}$                                                           & Approx.   & Data point & Privacy, Copyright, Robustness, Toxicity, Hallucination & M       & H     \\
KGA \citep{wang2023kga}                           & $A(D), D, D_f, D_{aux}$                                                        & Approx.   & Data point & General                                                  & M-L     & H     \\
\cite{eldan2023harry}                            & $A(D), D_f$                                                                       & Approx.   & Content    & Privacy, Copyright, Toxicity                             & H-M     & H     \\
Quark \citep{lu2022quark}                         & $A(D), D_{aux}, r(.)$                                                           & Approx.   & General    & General                                                  & H       & Sgnf  \\
PCGU \citep{yu2023unlearning}                     & $A(D), D_{aux}$                                                                 & Approx.   & Concept    & Fairness                                                 & H       & M     \\
Task vector \citep{ilharco2022editing}            & $A(D), D_f$                                                                       & Approx.   & Task       & Model capability                                         & H-M     & H     \\
F-Learning \citep{ni2023forgetting}               & $A(D), D_f$                                                                       & Approx.   & Task       & Model capability                                         & H-M     & H     \\
DEPN \citep{wu2023depn}                           & $A(D), D_f$                                                                       & Approx.   & Data point & Privacy                                                  & M-L     & M     \\
Selective pruning \citep{pochinkov2023dissecting} & $A(D), D_f$                                                                       & Approx.   & Task       & Model capability                                         & H       & M     \\
EUL \citep{chen2023unlearn}                       & $A(D), D_r, D_f$                                                                 & Weak      & Data point & Privacy, Copyright, Robustness, Toxicity                 & H-M     & M     \\
SISA-FC \citep{kumar2022privacy}                  & $A(D_s), D_{sr}, D_{sf}$                                                    & Weak/No   & Data point & Privacy, Copyright, Robustness, Toxicity                 & L       & L     \\
SISA-A \citep{kumar2022privacy}                   & $A(D_s), D_{sr}, D_{sf}$                                                    & Weak/No   & Data point & Privacy, Copyright, Robustness, Toxicity                 & M       & M     \\
LEACE \citep{belrose2023leace}                    & $A(D), D$                                                                          & Weak      & Concept    & Fairness                                                 & H-M     & M     \\
\cite{limisiewicz2022don}                        & $A(D), D_{aux}$                                                                 & Weak      & Concept    & Fairness                                                 & H-M     & L     \\
Zero-shot self-debiasing \citep{gallegos2024self} & $API(A(D)), D_f$                                                                  & Weak      & Concept    & Fairness                                                 & H       & L     \\
SERAC \citep{mitchell2022memory}                  & $API(A(D)), D_f$                                                                  & Weak      & General    & General                                                  & H       & M     \\
MemPrompt \citep{madaan2022memory}                & $API(A(D)), D_f$                                                                  & Weak      & General    & General                                                  & M       & L     \\
ICUL \citep{pawelczyk2023context}                 & $API(A(D)), D_f, D_{aux}$                                                      & Weak      & General    & General                                                  & H       & L     \\
\bottomrule
\end{tabular}}
}%
\end{adjustbox}
\end{table}

%Benet
%JOSEP. Changed title
\section{Evaluation of unlearning in LLMs}\label{section:evaluation}

%JOSEP. Rewritten.
In this section, we analyze the evaluation of forgetting in LLMs, including the datasets, models, and metrics employed by the surveyed works. The analysis of metrics focuses on the main aspects discussed in Section \ref{section:requirements}: i) whether the model has effectively forgotten the target knowledge, ii) whether the model retained the rest of its capabilities, and iii) the computational cost of the forgetting process. Hereafter, these measures will be referred 
%JOSEP. Replaced timeliness by runtime in all the paper.
to as forgetting, retaining, and runtime, respectively.

\subsection{Datasets}
\label{section:evaluation-datasets}

A proper assessment of forgetting generally requires three different datasets, which we refer to as \emph{forgetting training set}, \emph{forgetting test set}, and \emph{retaining test set}. The two forgetting sets 
%JOSEP. rewritten rest of paragraph.
are the unlearning counterparts of the training and test sets employed in traditional ML: the forgetting training set comprises samples representing the knowledge that the model should forget, whereas the forgetting test set facilitates the measurement of the forgetting generalization. The retaining test set is typically disjoint from the forgetting sets and is used to assess the utility ---preservation of  capabilities--- of the unlearned model.

%JOSEP. Slight edits in this paragraphs.
\autoref{table:datasets-frg} lists the forgetting datasets used by the surveyed methods. The ``Forgetting target'' column specifies the forgetting request type (as defined in Section \ref{section:types}) and a more particular description of the undesired knowledge to be unlearned. The table shows that most works align with \emph{feature or concept removal} (such as problematic generations or biases) and \emph{item removal} ({\em e.g.}, a subset of samples) requests. Only \cite{pochinkov2023dissecting} are found to work on \emph{task removal}.
%Benet DOUBT: ¿El siguiente párrafo aporta? ¿Corresponde a este lugar?
%JOSEP. Suposo que no molesta.
It is important to note that some authors select combinations of undesired knowledge and forgetting request type that do not align with real-world scenarios, although this choice could be justified for experimental purposes. On one hand, \cite{wu2023depn} and \cite{borkar2023what} consider Personal Identifiable Information (PII) as items to be forgotten. Nevertheless, sensitive information should be treated as a \emph{concept} to be forgotten, to preclude the model from providing any details compromising privacy. For example, for an individual born in Paris, even if the model does not generate the city name verbatim, it can disclose that the birthplace is 'The city of light', sobriquet of Paris. On the other hand, \cite{eldan2023harry} experiment with forgetting a copyrighted corpus (Harry Potter books) as a concept. This may be excessive to avoid copyright laws, for which it is sufficient not to generate copies of the text. Forgetting copyrighted texts as \emph{items} should be enough for these cases.

%David: for consistency, maybe it would be nice to make this types explicit in the table. Also, what about task or class removal (there were some mentions to them in the previous section by Najeeb).
%Benet NOTE: Forgetting request types added to the table
%Benet DOUBT: Add a column with forgetting motivations? (e.g., alignment, copyright...). Some papers don't perfectly align, but maybe it's interesting
%Benet DOUBT: I added the example of task removal, but I didn't found an example of class removal already. Should I mention that there is no "class removal" method found for LLMs?
%Alberto: as I mentioned in the section on forgetting types, class removal is weird in generative LLMs, as it would refer to removing some token from the vocabulary. Class removal in sentiment analysis or similar classification tasks should be very similar in LLMs and other types of models.

The datasets in \autoref{table:datasets-frg} are split into the forgetting training and test sets. For example, \cite{yao2023large} use the training and test splits of PKU-SafeRLHF \citep{ji2023beavertails} to compare the rate of harmful answers for forgotten and unseen prompts. Authors focusing on forgetting a copyrighted corpus or specific samples ({\em e.g.}, data of an individual) do not use a forgetting test set \citep{wu2023depn, pawelczyk2023context}. Instead, they measure the forgetting success based on the model's inability to generate verbatim copies of the text to be forgotten or correctly process the samples whose forgetting is desired. This will be later expanded in Section \ref{section:evaluation-forgetting}.

\begin{table}[]
\small
\renewcommand{\baselinestretch}{1} 
\centering
\caption{Datasets used for forgetting}
\label{table:datasets-frg}
\begin{adjustbox}{angle=90}
\scalebox{0.85}{
\rowcolors{2}{gray!20}{gray!5}
\begin{tabular}{llp{7cm}} %"p{7cm}" used to avoid a too wide row caused by citations
\hline
Dataset name                                 & Forgetting target                  & Paper/s                     \\ \hline
PKU-SafeRLHF \citep{ji2023beavertails}    & Concept: Harmful generations         & \cite{yao2023large}         \\ %\hline
HaluEval \citep{li2023halueval}           & Concept: Hallucinated generations    & \cite{yao2023large}         \\ %\hline
RealToxicityPrompts \citep{gehman2020real}& Concept: Toxic generations           & \cite{lu2022quark,ilharco2022editing}          \\
Civil Comments \citep{borkan2019nuanced}  & Concept: Toxic generations           & \cite{ilharco2022editing} \\
Harry Potter Universe \citep{eldan2023harry}  & Concept: Related generations           & \cite{eldan2023harry}       \\ %\hline
WinoBias \citep{zhao2018gender}           & Concept: Gender bias                 & \cite{limisiewicz2022don}   \\ %\hline
WinoMT \citep{stanovsky2019evaluating}    & Concept: Gender bias                 & \cite{limisiewicz2022don}   \\ %\hline
CrowS Pairs \citep{nangia2020crows}       & Concept: Gender bias                 & \cite{yu2023unlearning}     \\ %\hline
Winogender Schemas \citep{rudinger2018gender}& Concept: Gender bias              & \cite{yu2023unlearning}     \\ %\hline
Bias in Bios \citep{dearteaga2019bias}    & Concept: Gender bias                 & \cite{belrose2023leace}      \\ %\hline
StereoSet \citep{nadeem2021stereoset}     & Concept: Stereotype bias            & \cite{yu2023unlearning}     \\ %\hline
English Universal Dependencies \citep{nivre2020universal} & Concept: Part-of-Speech & \cite{belrose2023leace}     \\ %\hline
RedPajama \citep{together2023redpajama}   & Concept: Part-of-Speech        & \cite{belrose2023leace}     \\ %\hline
Training Data Extraction Challenge \tablefootnote{\url{https://github.com/google-research/lm-extraction-benchmark}}& Item: Specific generations   & \cite{jang2022knowledge}    \\ %\hline %Subset in line with Pile \cite{gao2021pile} FORGET SPECIFIC SAMPLES OF 200 TOKENS
Pile \citep{gao2021pile}                  & Item: Samples $\And$ Concept: PoS     & \cite{jang2022knowledge, belrose2023leace,pochinkov2023dissecting}     \\ %\hline
Harry Potter and the Sorcerer’s Stone \citep{jk2002harry}  & Item: Copyright corpus  & \cite{yao2023large} \\
SST2 \citep{socher2013sst2}               & Item: Samples             & \cite{pawelczyk2023context} \\ %\hline
Amazon polarity \citep{zhang2015yelp}     & Item: Samples             & \cite{pawelczyk2023context} \\ %\hline
Yelp polarity \citep{zhang2015yelp}       & Item: Samples             & \cite{pawelczyk2023context} \\ %\hline
IMDB \citep{maas2011imdb}                 & Item: Samples             & \cite{chen2023unlearn}      \\ %\hline
SAMSum \citep{gliwa2019samsum}            & Item: Samples             & \cite{chen2023unlearn}      \\ %\hline
LEDGAR \citep{tuggener2020ledgar}         & Item: Samples             & \cite{wang2023kga}          \\ %\hline
PersonaChat \citep{zhang2018personalizing}& Item: Samples             & \cite{wang2023kga}          \\ %\hline
IWSLT14 \citep{cettolo2014iwslt}          & Item: Samples             & \cite{wang2023kga}          \\ %\hline
Enron emails \citep{klimt2004enron}       & Item: Private information        & \cite{wu2023depn,borkar2023what} \\ 
zsRE \citep{levy2017zero}                 & Item: Old facts                  & \cite{ni2023forgetting} \\
CounterFact \citep{meng2022locating}      & Item: Old facts                  & \cite{ni2023forgetting} \\
CodeParrot GitHub Code \citep{tunstall2022natural} & Task: Programming& \cite{pochinkov2023dissecting} \\
\hline
\end{tabular}
}
\end{adjustbox}
\end{table}

\autoref{table:datasets-rt} lists datasets used to assess the model's retaining. The ``Retaining target'' column specifies the preserved capability under evaluation. Note that the zsRE \citep{levy2017zero} dataset and those subsequently listed were also employed for forgetting purposes by the same authors (see \autoref{table:datasets-frg}). This is because the objective was to selectively forget a subset of samples, specific generations ({\em i.e.}, private information in Enron emails \citep{klimt2004enron}), a concept ({\em i.e.}, gender bias in (Bias in Bios \citep{dearteaga2019bias}) or a task ({\em e.g.}, Python programming in CodeParrot GitHub Code \citep{tunstall2022natural}), while retaining the rest dataset-related skills. Several researchers choose widely recognized benchmarks for LLMs \citep{eldan2023harry,jang2022knowledge,yao2023large,pawelczyk2023context,chen2023unlearn} to evaluate the preservation of overall capabilities. In contrast, a limited number of authors employ datasets with tasks that are closely aligned with the knowledge intended for forgetting \citep{limisiewicz2022don,belrose2023leace,pochinkov2023dissecting,ni2023forgetting}. As will be detailed in Section \ref{section:evaluation-retaining}, this strategy is taken because tasks that are similar but peripheral to the focus of the forgetting process are expected to be the most affected by it.
%Benet DOUBT: Tal vez hay demasiado overlap entre lo que se dice aquí (e.g., datasets generales y específicos) y en la sección de "Retaining evaluation"?

\begin{table}[]
\renewcommand{\baselinestretch}{1} 
\small
\centering
\caption{Datasets only used for retaining}
\begin{adjustbox}{angle=90}
\label{table:datasets-rt}
\scalebox{0.9}{
\rowcolors{2}{gray!20}{gray!5}
\begin{tabular}{llp{7cm}} %"p{7cm}" used to avoid a too wide row caused by citations
\hline
Dataset name                             & Retaining target            & Paper/s                     \\ \hline
Wizard of Wikipedia \citep{dinan2018wizard}& Dialogue                  & \cite{jang2022knowledge}     \\ %\hline
Empathetic Dialogues \citep{rashkin2019towards}& Dialogue              & \cite{jang2022knowledge}     \\ %\hline
Blended Skill Talk \citep{smith2020put}   & Dialogue                   & \cite{jang2022knowledge}     \\ %\hline
Wizard of Internet \citep{komeili2022internet}& Dialogue               & \cite{jang2022knowledge}     \\ %\hline
piqa \citep{bisk2020piqa}                 & Q\&A                       & \cite{eldan2023harry, jang2022knowledge}       \\ %\hline
COPA \citep{gordon2012COPA}               & Q\&A                       & \cite{jang2022knowledge}     \\ %\hline
ARC \citep{clark2018arc}                  & Q\&A                       & \cite{jang2022knowledge}     \\ %\hline
MathQA \citep{amini2019mathqa}            & Q\&A                       & \cite{jang2022knowledge}     \\ %\hline
PubmedQA \citep{jin2019pubmedqa}          & Q\&A                       & \cite{jang2022knowledge}     \\ %\hline
TruthfulQA \citep{lin2022truthfulqa}      & Q\&A                       & \cite{yao2023large}         \\ %\hline
GAP Coreference \citep{webster2018mind}   & Coreference                & \cite{limisiewicz2022don}   \\ %\hline % Coreference with bias
English Web Treebank \citep{silveira2014gold}& Dependency              & \cite{limisiewicz2022don}   \\ %\hline
WinoGrande \citep{sakaguchi2020winogrande}& Completition               & \cite{eldan2023harry,jang2022knowledge}\\
HellaSwag \citep{zellers2019hellaswag}    & Completition               & \cite{eldan2023harry,jang2022knowledge}\\ 
Lambada \citep{paperno2016lambada}        & Completition               & \cite{jang2022knowledge}    \\ %\hline
zsRE \citep{levy2017zero}                 & Completition               & \cite{ni2023forgetting} \\
CounterFact \citep{meng2022locating}      & Completition               & \cite{ni2023forgetting} \\
Pile \citep{gao2021pile}                  & Samples subset             & \cite{jang2022knowledge,pochinkov2023dissecting}     \\
SST2 \citep{socher2013sst2}               & Samples subset             & \cite{pawelczyk2023context} \\ %\hline
Amazon polarity \citep{zhang2015yelp}     & Samples subset             & \cite{pawelczyk2023context} \\ %\hline
Yelp polarity \citep{zhang2015yelp}       & Samples subset             & \cite{pawelczyk2023context} \\ %\hline
IMDB \citep{maas2011imdb}                 & Samples subset             & \cite{chen2023unlearn}      \\ %\hline
SAMSum \citep{gliwa2019samsum}            & Samples subset             & \cite{chen2023unlearn}      \\ %\hline
LEDGAR \citep{tuggener2020ledgar}         & Samples subset             & \cite{wang2023kga}          \\ %\hline
PersonaChat \citep{zhang2018personalizing}& Samples subset             & \cite{wang2023kga}          \\ %\hline
IWSLT14 \citep{cettolo2014iwslt}          & Samples subset             & \cite{wang2023kga}          \\ %\hline
Enron emails \citep{klimt2004enron}       & Generation                 & \cite{wu2023depn,borkar2023what} \\
CodeParrot GitHub Code \citep{tunstall2022natural} & Generation        & \cite{pochinkov2023dissecting} \\
WikiText-103 \citep{merity2016pointer}    & Generation                 & \cite{ilharco2022editing} \\
Bias in Bios \citep{dearteaga2019bias}    & Classification             & \cite{belrose2023leace} \\ \hline %Gender-profession classification
\end{tabular}
}
\end{adjustbox}
\end{table}

\subsection{Models}
\label{section:evaluation-models}

\autoref{table:models} reports the LLMs used by each of the surveyed methods.
Models are sorted by their number of parameters, differing by as much as 3 orders of magnitude ({\em i.e.}, from 11 million to 30 billion parameters). 
LLMs are often released in various sizes to accommodate a wide range of use cases. The table enumerates the specific sizes used, separated by commas. In cases where multiple authors selected the same model but in different sizes, a row without the model name is used. As suggested by \cite{carlini2021extracting}, larger models tend to memorize more, making them more challenging and interesting for evaluation. Notably, 8 works opted for models with over 1 billion parameters.

\begin{table}[]
\small
\renewcommand{\baselinestretch}{.8} 
\caption{LLMs used for forgetting}
\label{table:models}
\centering
\begin{adjustbox}{angle=90}
\rowcolors{2}{gray!20}{gray!5}
\begin{tabular}{llp{7cm}} %"p{7cm}" used to avoid a too wide row caused by citations
\hline
\multicolumn{1}{l}{Model name}                   & Number of parameters & Paper/s                          \\ \hline
\multicolumn{1}{l}{ALBERT \citep{lan2020albert}}      & 11M           & \cite{yu2023unlearning}          \\ %\hline
\multicolumn{1}{l}{DistilBERT \citep{sanh2019distilbert}}      & 67M           & \cite{wang2023kga}          \\ %\hline
\multicolumn{1}{l}{BERT \citep{devlin2019bert}}      & 110M           & \cite{wu2023depn, yu2023unlearning,limisiewicz2022don,belrose2023leace}          \\ %\hline
\multicolumn{1}{l}{ELECTRA \citep{clark2020electra}}    & 110M           & \cite{limisiewicz2022don}          \\ %\hline
\multicolumn{1}{l}{RoBERTa \citep{liu2019roberta}}       & 355M           & \cite{pochinkov2023dissecting} \\ 
                                                        & 125M           & \cite{yu2023unlearning} \\
\multicolumn{1}{l}{GPT-2 \citep{radford2019language}}    & 117M, 355M, 774M      & \cite{ilharco2022editing}\\
                                                        & 355M, 774M            & \cite{lu2022quark}\\
\multicolumn{1}{l}{Bloom \citep{workshop2022bloom}}      & 560M, 1.1B            & \cite{pawelczyk2023context}      \\ 
\multicolumn{1}{l}{Phi \citep{li2023textbooks}}          & 1.5B                  & \cite{eldan2023harry}            \\ 
\multicolumn{1}{l}{GPT-Neo \citep{black2021gpt}}         & 125M, 1.3B, 2.7B      & \cite{jang2022knowledge}         \\ 
\multicolumn{1}{l}{T5 \citep{raffel2020exploring}}       & 223M, 3B              & \cite{chen2023unlearn}           \\ 
\multicolumn{1}{l}{OPT \citep{zhang2022opt}}             & 125M, 1.3B, 6.7B      & \cite{pochinkov2023dissecting}\\
                                                        & 125M, 1.3B, 2.7B      & \cite{jang2022knowledge}\\
                                                        & 1.3B, 2.7B            & \cite{yao2023large}\\
\multicolumn{1}{l}{Galactica \citep{taylor2022galactica}}& 125M, 1.3B, 6.7B      & \cite{pochinkov2023dissecting}\\
\multicolumn{1}{l}{Llama-2 \citep{touvron2023llama2}}    & 7B                    & \cite{yao2023large,ni2023forgetting} \\ 
\multicolumn{1}{l}{Pythia \citep{biderman2023pythia}}    & 160M, 1.4B, 6.9B, 12B & \cite{belrose2023leace}          \\ 
                                                         & 160M, 1.4B, 6.9B      & \cite{pochinkov2023dissecting} \\
\multicolumn{1}{l}{Llama \citep{touvron2023llama1}}      & 7B, 13B, 30B          & \cite{belrose2023leace}  \\
                                                         & 7B                    & \cite{eldan2023harry} \\
\hline
\end{tabular}
\end{adjustbox}
\end{table}

%JOSEP. Changed title.
\subsection{Metrics and attacks to evaluate forgetting}
\label{section:evaluation-forgetting}
Most surveyed methods only offer approximate unlearning and, therefore, do not provide forgetting guarantees. In these cases, the forgetting success should be empirically evaluated \emph{ex post}.

Forgetting success is usually assessed by quantifying the decrease in the level of undesired knowledge from before to after the forgetting process. By undesired knowledge level, we mean the model's capability/likelihood of making the kind of predictions that the model owners aim to forget ({\em e.g.}, harmful, copyrighted, or privacy-leaking generations).
Specifically, the methods aim to minimize the undesired knowledge level in both the forgetting training and test sets. A reduction in the forgetting training set implies that seen samples are being forgotten, whereas a reduction in the forgetting test set indicates that forgetting is affecting even unseen samples ({\em i.e.}, it is able to generalize). If the observed decrease in the training set is much more significant than in the test set, the forgetting approach is assumed to ``overfit'' the training samples. Considering the capabilities of LLMs for memorizing text \citep{carlini2021extracting}, it is possible that the forgetting process only worked for samples in the forgetting training set, and even a slight paraphrasing could be sufficient to obtain undesired predictions.
% Benet DOUBT: ¿Definir "undesired knowledge" al inicio de la sección? Se usa una vez en Retaining y puede que estubiera bien usarlo en Datasets.
%David: puede valer, pero entonces estaria bien definirlo antes, por ejemplo, en el background, cuando se habla de forgetting requests -> olvidar undesired knowledge (feature/behaviors/class, etc.). 
%Benet DOUBT: Mencionado (puedes buscar undesired knowledge para verlo). Haría falta detallarlo más, o el nombre es autodefinitorio?

In the following, we examine the metrics employed by the surveyed works to measure the attained level of forgetting:
\begin{itemize}
    \item \textbf{Dataset-specific metrics}: Most papers \citep{eldan2023harry,jang2022knowledge,yao2023large,pawelczyk2023context,chen2023unlearn,ni2023forgetting,pochinkov2023dissecting} leverage the predefined performance metric of the forgetting dataset, that is, the accuracy in a classification or completion dataset.

    \item \textbf{Toxicity}: Papers seeking to mitigate toxicity of generated texts \citep{ilharco2022editing,lu2022quark} sometimes rely on off-the-shelf metrics, such as Perspective API\footnote{\url{https://github.com/conversationai/perspectiveapi}} or Toxicity\footnote{\url{https://github.com/unitaryai/Toxicity}}. These metrics involve training an LLM on a toxicity dataset to predict the toxicity score of input text.
    
    \item \textbf{Bias}: Works addressing bias in LLMs ({\em e.g.}, gender, race or stereotype) \citep{belrose2023leace,limisiewicz2022don,yu2023unlearning} commonly rely on the bias-sensitive prediction probability. For example, for pronoun prediction of a person with a known occupation ({\em e.g.}, doctor), probabilities for both pronouns are expected to be 50\%. Under the premise of profession-induced gender bias, \cite{dearteaga2019bias} introduced the TPR-GAP metric, which assesses the difference (GAP) in the true positive rate (TPR) between the two genders for each occupation. On the same basis, \cite{limisiewicz2022don} presented the relative gender preference (RGP) metric, which can be roughly defined as the difference in gender probabilities for texts with and without the profession name.
    
    \item \textbf{Generation}: For undesired knowledge without any specific metrics, evaluation often focuses on how easily the model can generate that undesired knowledge. This differs depending on whether the forgetting process aims to avoid generating exact or similar text to that of the forgetting request:
    \begin{itemize}
        \item \textit{Exact generation}: Here the focus is on preventing the model from generating verbatim copies of the text to be forgotten. This is common in scenarios involving copyright, where we do not want the model generating a text in the same way as the source. Most authors consider that such exact reproductions will probably be the result of feeding the model with that undesirable text. On this basis, the perplexity metric (which is standard in text generation) is often employed \citep{jang2022knowledge,yao2023large,wu2023depn,pochinkov2023dissecting}. This approach is sometimes followed when the objective is to avoid the generation of Personal Identifiable Information (PII). \cite{jang2022knowledge} propose the extraction likelihood metric, which considers overlaps of n-grams, and the memorization accuracy metric, for matching of identical token sequences. They define unlearning as complete when both metrics are lower than a threshold (akin to a privacy guarantee) for the forgetting test set. \cite{wu2023depn} adapt their metrics to the type of PII to forget, such as exposure for phone numbers and mean reciprocal rank for names. Note that most of these evaluations do not use a forgetting test set, but focus on \textit{exact} replicates, which are the consequence of using the forgetting training set as input. This seems an insufficient approach for PII (or any other private information) since the target for privacy preservation should be to avoid the generation of the targeted sensitive concept, not to prevent a concrete way of expressing it (unlike copyright).

      %JOSEP. A bit rewritten.
        \item \textit{Similar generation}: Here any generations are considered that indicate that the model has been trained with undesired knowledge. For instance, \cite{wang2023kga} measure output similarity with the text to be forgotten by using the Jensen-Shannon Divergence (JSD), the Language model Probability Distance (LPD), and the Proportion of instances with Decreased Language model Probability (PDLP), while \cite{yao2023large}
        %JOSEP. Added meaning of BLEU
        leverage the BiLingual Evaluation Understudy (BLEU) score. \cite{eldan2023harry}, who aimed to forget the Harry Potter universe corpora, opt for a more fine-grained evaluation by using a forgetting test set specifically curated to elicit related knowledge ({\em e.g.}, ``When Harry returned to class, he observed his best friends,'').
    \end{itemize}    
    
    \item \textbf{Membership Inference Attack (MIA)}: A subset of authors \citep{wang2023kga,chen2023unlearn,pawelczyk2023context} use the success of a membership inference attack as a metric. This type of attack aims to determine whether a piece of knowledge was used during the (pre-)training of the model.    
    See Section \ref{section:motivations-privacy} for a brief introduction to privacy issues and the role of MIA, and Section \ref{sec:mia} for more details on specific attacks and definitions.
    This inversely aligns with the forgetting objective of nullifying the influence of undesired knowledge on the model (thereby being untraceable by MIAs).
    %David: since Alberto's text on attacks is quite long, just mention the use of attacks as metrics here and refer to the dedicated section for more details.
    %Alberto: extended to take into account the subsection on MIAs
\end{itemize}

% Starting scenarios
Two different scenarios are commonly employed as starting points for forgetting, one of them being more realistic and the other one more controlled. In the more realistic scenario \citep{jang2022knowledge,yao2023large,yu2023unlearning,limisiewicz2022don}, forgetting is directly applied to a standard (usually publicly available) LLM. The undesired knowledge is then assumed to be somehow part of the pre-training data. For example, most LLM pre-training datasets are assumed to contain hard-to-filter stereotypical biases inherent to the human-written text employed for training. This scenario imposes restrictions on the kind of knowledge to be forgotten: i) undesired knowledge must have sufficient presence in the pre-training dataset for the LLM to learn it, and ii) if the pre-training data samples related to the undesired knowledge are unknown (as it is typical in the case of biases), an external dataset is required as a representation of that knowledge because forgetting is data-driven. 

These constraints can complicate the exploration for certain forgetting tasks, leading some authors \citep{eldan2023harry,chen2023unlearn,borkar2023what,wu2023depn,wang2023kga} to opt for a second less realistic but more controllable scenario. In this case, forgetting is applied to an LLM fine-tuned with the (dataset representing the) undesired knowledge. This way, researchers can experiment with any learnable domain while having significant control over the undesired knowledge. That control can be absolute if that knowledge set is completely disjoint from the pre-training dataset ({\em e.g.}, very specific or not publicly available). Moreover, the original LLM can be used as the forgetting ground truth, by assuming it has (presumably) never been trained on the undesirable knowledge. In contrast, in the realistic scenario, the authors requiring a ground truth need to re-train the LLM from scratch with a version of the pre-training dataset that does not include the undesired knowledge. The lack of realism in the controlled scenario arises from two primary factors. First, undesired knowledge is the last to be learned by the model, making it possible to overfit to it. This may cause the undesired knowledge to be prominent in the model, thereby resulting in an excessively high initial undesired knowledge level and an easier (or at least unrealistic) forgetting process. The sources of the undesired knowledge within the (typically Gargantuan) pre-training dataset are usually unknown. And what is known is just that the model has that undesired behavior for very specific cases, and that undesired learning may have occurred at any time during pre-training.

%Benet DOUBT: El concepto de los onion points que se menciona en el siguiente pàrrafo comentado, ¿es suficientemente interesante para ser comentado? ¿Deberia ser comentado aquí o mejor en una sección más introductoria para remarcar la complejidad del problema?
%"We embed email addresses from the Enron dataset6 in the WikiText-2 dataset (Merity et al., 2016) and fine-tune GPT-2 small on it (mod; dat). Initially, the dataset had 6523 email addresses and we were able to extract 44 out of them. We unlearn these 44 email addresses by removing them from the dataset7 and fine-tuned GPT-2 on the unlearned dataset. After unlearning, we found that 20 new email addresses got leaked by the model which were previously safe. We call them \textit{onion points}"

%Alberto: provisional placing of this section
%David: I skip the revision of this part until properly integrated
%Benet: Alberto, en un comentario previo dice que "since Alberto's text on attacks is quite long, just mention the use of attacks as metrics here and refer to the dedicated section for more details.", así que lo dejaría dónde estaba y lo referencio.
\subsubsection{Membership Inference Attacks in LLMs}
\label{sec:mia}
\cite{carlini2021extracting} propose a series of black-box data extraction attacks
on pre-trained LLMs based on membership inference. They conducted attacks on GPT-2,
which is a closed model (both in parameters and training data), but had limited access to the GPT-2
training data (through colleagues) to verify their findings.
Their proposed baseline attack is to generate 200,000 samples of unconditioned text with a
generative model, beginning each sample with a special \textit{beginning-of-sequence} token.
Sequences are generated using a top-$k$ sampling strategy until an \textit{ending-of-sequence} token
is reached. The samples are then ranked based on the perplexity assigned to each sequence
by the LLM.

The attack is then refined in two ways, by changing the sequence generation process and by
focusing on different metrics.
Regarding sample generation, the authors propose two refinements.
The first one consists in changing the temperature during sequence generation, starting with
a high temperature and reducing it progressively until it reaches a value of $1$. This allows
the LLM to first explore different paths in the generation of the first tokens, and then follow
a clearer sequence of (possibly) seen data. Their second approach consists in conditioning the
text generation with data from the internet. Note that LLMs are pre-trained on big datasets from
publicly accessible data on the internet and thus using data from the same sources could mean
that there are intersections between the training dataset and the attacker knowledge.
Regarding metrics, the authors use the following ones: perplexity as in the baseline attack,
the ratio of the log-perplexity assigned to the sample by the generating model over the log-perplexity 
assigned to the sample by a different model (mimicking the shadow-model approach by~\cite{shokri2017membership}),
%JOSEP. Rewritten.
the ratio of the log-perplexity over the zlib compression library,
the entropy of the sample, the ratio of perplexities between the generated sample and a lowercased
version of the same sample, and the minimum perplexity across any 50-token long sliding windows.

The authors show that exact sequences from the training data are returned from the model, 
including personal data such as names and contact information, copyrighted material such as
news pieces, and potentially security-related information, such as logs and configuration 
files. 
The work suggests that bigger models tend to leak more exact sequences.
% Table~\ref{tab:carlini_attack_extracted_data} shows the categories of extracted data, along
% with the counts for each category. The table is quoted directly from \cite{carlini2021extracting}.

% \begin{table}
% \small
% \renewcommand{\baselinestretch}{1} 
%     \centering
%     \begin{tabular}{lc}
%         \hline
%         \textbf{Category} & \textbf{Count} \\
%         \hline
%         US and international news & 109 \\
%         Log files and error reports & 79 \\
%         License, terms of use, copyright notices & 54 \\
%         Lists of named items (games, countries, etc.) & 54 \\
%         Forum or Wiki entry & 53 \\
%         Valid URLs & 50 \\
%         Named individuals (non-news samples only) & 46 \\
%         Promotional content (products, subscriptions, etc.) & 45 \\
%         High entropy (UUIDs, base64 data) & 35 \\
%         Contact info (address, email, phone, twitter, etc.) & 32 \\
%         Code & 31 \\
%         Configuration files & 30 \\
%         Religious texts & 25 \\
%         Pseudonyms & 15 \\
%         Donald Trump tweets and quotes & 12 \\
%         Web forms (menu items, instructions, etc.) & 11 \\
%         Tech news & 11 \\
%         Lists of numbers (dates, sequences, etc.) & 10 \\
%         \hline
%     \end{tabular}
%     \caption{Categories of memorized data extracted by the attack.
%     Table quoted verbatim from~\cite{carlini2021extracting}.
%     Besides personal data, such as names or contact information, content potentially subjected to
%     copyright was extracted, such as news pieces.}
%     \label{tab:carlini_attack_extracted_data}
% \end{table}

The authors provide the $k$-eidetic memorization privacy/security criteria, defined as follows.

\begin{definition}
\label{def:knowl_extraction}
\textbf{Model knowledge extraction.} A string $s$ is extractable from an LLM $f_\theta$ if there
exists a prefix $c$ such that
$s \leftarrow \argmax_{s':|s'|=N} f_\theta(s'|c). $
\end{definition}

\begin{definition}
\label{def:eidetic_memo}
\textbf{$k$-Eidetic memorization.} A string $s$ is $k$-eidetic memorized (for $k \ge 1$)
by an LLM $f_\theta$ if $s$ is extractable from $f_\theta$ and $s$ appears in at most $k$ examples
%JOSEP. A bit rewritten.
in the training data $X$, that is $|\{x \in X: s \subseteq x\}| \le k$.
\end{definition}

Thus, two limitations on memorization are introduced.
The first is the length of the memorized strings.
The second is the number $k$ of times that a given string appears in the training data. 

\cite{nasr2023scalable} follow up on the work by \cite{carlini2021extracting} by executing attacks
on open-source and semi-open large language models. The key difference in this work is that the
adversaries have access to the training dataset or at least enough information to build a similar
dataset.
The authors confirm the findings of \cite{carlini2021extracting}, including the observation
that larger and more capable models are more vulnerable to data extraction attacks.
Then, they conduct attacks on ChatGPT (with gpt-3.5-turbo) and find that the same approach
does not make the models leak any information.

The authors suggest that fine-tuning --including RLHF-- to make the models act as conversational
agents prevents them from reverting to training data from their pre-training phase. 
However, the authors launch a \textit{divergence attack}.
As mentioned above, generative models sample from the distribution of tokens conditioned
on the previous tokens. In their attack, they make ChatGPT repeat a single token, expecting that
the generation will diverge from the expected path. In the example provided in the paper, the
authors make ChatGPT repeat the sequence ``poem, poem, poem...'' indefinitely. Even if the most
probable token at each generation step is the word ``poem'', nonzero probabilities are assigned
to other tokens. Sampling repeatedly from the same distribution may cause the generator to output
a low-probability token, thus diverging from the expected path. Once the generation diverges,
ChatGPT seems to fall back to memorized data. By comparing the behavior of gpt-3.5-turbo and
gpt-3.5-instruct-turbo, which are (presumably) fine-tuned on different data, the authors conclude
that the leaked data comes from the data used during model pre-training, which likely includes
personal, sensitive, and copyrighted material. Namely, the authors recover personal data,
inappropriate content, fragments of books, URLs, UUIDs, code snippets, research papers, boilerplate
texts, and mixed memorized data.

The authors provide two security/privacy criteria.
%JOSEP. I uniformize the name of the training set to X, as in previous definitions.
\begin{definition}
\label{def:extractable_memo}
\textbf{Extractable memorization.} Given a generative large language model $Gen$, an example $s$ from the training set $X$ is extractably memorized if an adversary (without access to $X$) can construct a prompt $c$ that makes the model produce $s$ (i.e., $Gen(c) = s$).
\end{definition}

\begin{definition}
\label{def:discoverable_memo}
\textbf{Discoverable memorization.} For a model $Gen$ and an example $[c||s]$ from the training set $X$, we say that $s$ is discoverably memorized if $Gen(c) = s$.
\end{definition}

\cite{patil2023can} present two white-box access data recovery attacks.
In this case, the authors attempt to extract information from the models after some
digital forgetting procedure has been applied to them.

In this paper, similarly to previous works, the objective of the attacker is to obtain
some answer $s$ to a question $c$, where the pair $(c,s)$ is sensitive information. 
However, the authors present several threat models defined by how many attempts the
attacker has. In particular, the attacker is successful if the answer $s$ is within
a set of candidate answers $\mathcal{S}$. 
%JOSEP. Rewritten
The size $|\mathcal{S}|=B$ of the candidate set is defined as the attack budget.

The attack scenarios are as follows:
\begin{enumerate}
    \item \textit{Password attempts}. The attacker does not know $s$ but can verify its correctness in $B$ attempts. This represents an attacker trying to recover a password from a model to steal a personal account.
    \item \textit{Parallel pursuit}. Again, the attacker does not know $s$ \textit{ex ante} but can use the information in $S$ in parallel (and does not care if some candidates are incorrect). An example of this scenario is an attacker building an address book from leaked email addresses to conduct spam operations.
    \item \textit{Verification by data owner}. The attacker is the owner of datapoint $s$ and wants it to be deleted from the model. 
\end{enumerate}

The authors propose two white-box attacks based on the logit lens
%JOSEP. Unneeded footnote, because there is the reference.
%\footnote{Nostalgebrist, Interpreting GPT:
%the logit lens, LessWrong, 2020. \url{https://www.lesswrong.com/posts/AcKRB8wDpdaN6v6ru/interpreting-gpt-the-logit-lens}}
~\citep{nostalgebraist2020interpreting}
The logit lens is an interpretability tool that computes the product of the intermediate activations 
in (decoder-only) pre-trained transformers with the embedding layer, obtaining intermediate distributions
over the vocabulary. The logit lens can be used to explore how the prediction of a model evolves after
each intermediate block.

Their first attack is the Head Projection attack. It consists of obtaining the logit lens distribution of every layer in the LLM and taking the top-$k$ candidates in each distribution. The candidate set $S$ is the union of these candidates. 
Their second attack leverages changes in the probability assigned to each token
and is called the Probability Delta attack.
%JOSEP. Rewritten.
As in the previous attack, first the logit lens distributions of every layer are obtained.
Then, the differences in the distributions between every two consecutive layers are obtained.
Finally, the top-$k$ and bottom-$k$ candidates from these differences are included 
in the candidate list $S$.

The authors define their attack success metric as follows.

\begin{definition}
    \label{def:attack_success}
    \textbf{Attack Success Metric.} Given datapoints ${c_i,s_i}_{i=1}^N$: 
    $$ AttackSuccess@B(\mathcal{M}) = \frac{1}{N} \sum_{i=1}^N \mathds{1}[s_i \in S_i]$$
    where $S_i$ is the candidate set produced for model $\mathcal{M}$ on datapoint $c_i$ (with $|S_i| = B$),
    and $\mathds{1}[\cdot]$ is the indicator function.
\end{definition}

Note that this definition integrates concepts from Definition~\ref{def:discoverable_memo}, in which both the question and answers are known, but accounts for multiple attempts at extraction.

%David: I continue with the revision here
\subsection{Retaining evaluation}
\label{section:evaluation-retaining}

%JOSEP. Change performance to utility
The assessment of model retaining often involves evaluating the utility difference of the LLM before and after the forgetting process on tasks different from the undesired knowledge. The smaller the utility difference, the better the retention. Ideally, retention evaluation should encompass all possible knowledge except that to be forgotten. Nonetheless, this is unfeasible in most cases, given the wide range of capabilities that LLMs can learn. Instead, researchers define the forgetting test set by selecting the subset of the most relevant tasks for which utility should be retained.

In the following, we list the most prominent strategies used to measure retained utility:
\begin{itemize}

\item \textbf{General benchmarks}. \cite{eldan2023harry}, \cite{wang2023kga}, and \cite{jang2022knowledge} leverage well-established LLM benchmarks, such as those used for comparing LLMs \citep{sakaguchi2020winogrande,zellers2019hellaswag,paperno2016lambada,bisk2020piqa}. These benchmarks assess the model's capabilities in natural language understanding, reasoning, and human-like generation. In other words, this strategy aims to measure the preservation of the model's general capabilities. The utility on these general tasks is compared to find the model's weaknesses and strengths.
%JOSEP. Deleted sentence.
%which are critical for the final users' model selection.
%David: and what is the goal of comparing LLMs? Which are the principles of those benchmarks?
%Benet DOUBT: Done just above. ¿A citation is needed for the "comparison" part? Like Google's Gemini's paper (https://arxiv.org/abs/2312.11805) using human exams to compare performance with ChatGPT

% Performance loss
%eldan2023harry | [1,3]%, generally 2%
%wang2023kga | Dataset metrics = LEDGAR and IWSLT <1%, PersonaChat 6%
%jang2022knowledge | Classification(ACC) [1,2]%, generally 1% , Dialogue(F1) [1,8]% higher losses for smaller models

%JOSEP. A bit rewritten.
\item \textbf{Related but different}. A reasonable assumption is that knowledge conceptually close to that to be forgotten will be the first and most affected by the forgetting process. 
%JOSEP. performance -> utility
Therefore, the (declining) utility for this knowledge may be an appropriate indicator of whether the model is retaining the remaining capabilities or forgetting is affecting more than just the undesired knowledge. Following this principle, several authors evaluate retaining on knowledge domains related but different to the undesired knowledge \citep{yao2023large, lu2022quark, limisiewicz2022don, belrose2023leace, chen2023unlearn, wang2023kga, pawelczyk2023context, wu2023depn,ni2023forgetting,pochinkov2023dissecting}. The specific strategies employed differ in terms of closeness and/or relatedness to the forgetting target:

\begin{itemize}
\item \textit{Related dataset} \citep{lu2022quark,yao2023large,limisiewicz2022don,ilharco2022editing}: 
%JOSEP. Slightly rewritten.
These proposals select a dataset different from that used for forgetting, but such that it is more or less related to the undesired knowledge. For instance, \cite{lu2022quark} measured toxicity, fluency and diversity in the WritingPrompts dataset \citep{fan2018writing} to measure the retaining of the model's writing capabilities when forgetting toxic generations from the RealToxicPrompts dataset \citep{gehman2020real}. Similarly,  \cite{yao2023large} leveraged BLEURT \citep{sellam2020bleurt} and the \textit{deberta-v3-large-v2} reward model\footnote{\url{https://huggingface.co/OpenAssistant/reward-model-deberta-v3-large-v2}} to compare the generations for TruthfulQA \citep{lin2022truthfulqa} questions by the original LLM with those resulting from unlearning harmful Q\&A pairs of PKU-SafeRLHF \citep{ji2023beavertails}.
%David: esta bien en cualquier caso que el ejemplo sea de toxic generation, ya que es lo que más le interesa a Huawei
%Benet NOTE: Hecho

% Performance loss
% lu2022quark | Toxicity 60% improvement, Fluency 11% loss, Diversity 0%
% yao2023large | Reward {17%,18%,3%} of IMPROVEMENT, Similarity {21%,41%,213%} of loss
% ilharco2022editing | Perplexity 3%

\item \textit{Same dataset but different samples} \citep{chen2023unlearn,wang2023kga,pawelczyk2023context,ni2023forgetting,pochinkov2023dissecting}: Methods that aim to forget a subset of samples from a dataset often evaluate retaining on the remaining samples of that same dataset. For example, \cite{pochinkov2023dissecting} forget Python code samples from \cite{tunstall2022natural} while aiming to retain the remaining code samples.

% Performance loss
% chen2023unlearn | IMDB Test set [7,9.5]% T5-base [0.5,1]% T5-3B / Retained set [0,1]% generally 1%, SAMSum Test set [1,2]% T5-base [0,1]% T5-3B / Retained set [0,1]%
% pawelczyk2023context | [3,8]% for small model, [1,4]% for big model
% ni2023forgetting | [1,2]% , generally 1%
% pochinkov2023dissecting | AUC graphs, hard to extract a single loss
% wang2023kga | Already mentioned fo "General benchmarks"

%JOSEP. Slightly rewritten
\item \textit{Same dataset but different task} \citep{belrose2023leace,wu2023depn}: The forgetting (training and/or test) dataset is used to measure the model's capabilities in a task disjoint of that to be forgotten. For instance, the Bias in Bios dataset \citep{dearteaga2019bias} is employed by \cite{belrose2023leace} for both forgetting gender bias and measuring retaining in the job prediction task. Another example is provided by  \cite{wu2023depn}, who aim to forget 
personally identifiable information from the Enron emails dataset \citep{klimt2004enron}, and measure retaining as the generation quality for the rest of the corpus.

% Performance loss
% belrose2023leace | Profession-prediction accuracy from 79.3% to 77.3%, 2% loss
% wu2023depn | Perplexity from 3.07 to 3.11, 1% loss

\end{itemize}

\end{itemize}

%JOSEP. A bit rewritten
The utility degradation observed in these studies usually falls within the range of 1\% to 5\%, although it can occasionally reach up to 20\%. It has been noted by \cite{jang2022knowledge} and \cite{pawelczyk2023context} that larger models tend to exhibit better utility retention.
%David: igual que se ponen ejemplos arriba sobre escenarios de evaluación, igual valdria la pena reportar alguna de las diferencias de rendimiento reportadas, para hacerse una idea de cuánto rendimiento se puede perder o a cuánto se puede aspirar.
%Benet NOTE: Hecho en el párrafo de justo encima. Para cada categoria, tienes un comentario con los "Performance loss" de cada método. Como no he detectado un patrón para cada categoria, he hecho el párrafo resumen.

%David: Lo de 'speed' queda un poco raro. A decidir la nomenclatura a usar para referirse al coste/tiempo de ejecución del forgetting según como quede la sección de Requirements (Alberto)
%Benet NOTE: Renombrado a Timeliness, en referencia al requirement "Timeliness and scalability"
%JOSEP. Changed timeliness to runtime. 
\subsection{Runtime evaluation}
\label{evaluation-timeliness}

%JOSEP. A bit rewritten.
The runtime of forgetting should always be significantly lower than that required for retraining the LLM without the undesired knowledge. Such retraining from scratch is the straightforward but resource-intensive process 
whose avoidance is sought by the surveyed methods. However, forgetting is rarely a ``fast'' procedure, since it usually involves costly training-related steps such as fine-tuning.
%Moreover, as mentioned in Section \ref{section:requirements}, execution runtime can be critical in certain cases. For instance, in scenarios where forgetting requests are frequently received, such as when numerous individuals claim their Right To Be Forgotten.
%Benet DOUBT: David, you hid ^this^ (I don't know why), but Alberto mentions something in line in the Requirements/Timeliness part. Maybe it can be discommended.

\cite{wang2023kga,chen2023unlearn,limisiewicz2022don,jang2022knowledge,yao2023large,lu2022quark,yu2023unlearning} explicitly report the runtime of their methods. Values range from minutes \citep{wang2023kga,yao2023large,limisiewicz2022don,jang2022knowledge, yu2023unlearning} to hours \citep{chen2023unlearn} to even more than a day \citep{lu2022quark}.
However, directly comparing the runtimes reported by different works would be unfair, due to the different hardware configurations, models, and/or undesired knowledge.
%Benet DOUBT: Incluir "In any case, methods working at prediction time such as \cite{pawelczyk2023context}, are assumed to be the fastest."?

There are some workarounds to compensate for hardware discrepancies, although they are not universally applicable. Firstly, runtimes can be normalized to the most common GPU, adjusting runtimes from other GPUs based on their relative performance in deep learning, as assessed through specific benchmarks. However, the substantial memory requirements of LLMs often require multi-GPU configurations with the corresponding interconnections. The latency of these interconnections also influences runtimes, but this is rarely reported in the literature.
Secondly, for those methods whose temporal cost is predominantly determined by fine-tuning, the number of epochs can be used as a more general indicator of cost. Nonetheless, this information is only reported by a subset of works \citep{eldan2023harry,chen2023unlearn,jang2022knowledge,yao2023large,yu2023unlearning,wu2023depn,ni2023forgetting,ilharco2022editing} and can only be used when comparing identical models, data and forgetting tasks, which is rarely the case.
%David: sólo para confirmar. Entonces se entiende que sólo se mide el coste en terminos de runtime y/o epochs.
%Benet NOTE: Creo que todos excepto el que reporta costes teóricos temporales y de memoria.

\section{Challenges and potential solutions}
\label{sec:challenges}

Section~\ref{sec:bgforgetting} discussed the motivations, types, and requirements of digital forgetting.
In this section, we evaluate the surveyed methods in terms of the presented requirements, focusing on
the provided forgetting guarantees, the performance of the resulting models, and the computational costs
and scalability of the methods.

\subsection{Guarantees of forgetting}

%Alberto: forgetting guarantees
%JOSEP. A bit rewritten.
The forgetting guarantees presented in Section~\ref{section:requirements} were mainly devised for general neural network classifiers.
Although they can still be used for LLMs, some factors limit their applicability.
On one hand, the guarantees refer to the parameters of the resulting models after training or forgetting.
That means that, for an \textit{ex ante} guarantee of forgetting, a clear approximation of the influence of every training sample on the model weights is known or can be computed.
While this may be feasible for small or simpler models, the complexity of LLMs may make such analyses difficult if not impossible.
% Najeeb: Challenges of several approaches

% Najeeb: Difficulty in identifying unlearning knowledge. 
%JOSEP. Moved here and slightly rewritten.
On the other hand, while unlearning an image or a record is relatively straightforward in typical machine unlearning, the complexity increases when dealing with text data in LLMs.
If the goal is to unlearn a text sequence containing sensitive information, not only is identifying explicit 
%JOSEP. A bit rewritten.
sensitive data difficult, but identifying implicit 
sensitive data (data that may allow sensitive inferences) is even more challenging. 
This challenge also applies when unlearning copyrighted documents, especially in non-fiction texts, where identifying the unique tokens associated with these documents becomes tedious.
When the aim is to eliminate model bias or hallucination, the complexity increases further. 
%JOSEP. Rewritten.
Bias is often widespread and difficult to detect in training text data, as it appears in scattered patterns across many examples in both explicit and implicit forms. 
Thus, unlearners often resort to identifying biased concepts within internal model representations and addressing them at this level.
Similarly, addressing hallucinations poses a significant challenge, as they often stem from multiple sources, making identification a nontrivial task.

%Najeeb: Idea forgetting approach.
%JOSEP. Moved here.
Furthermore, methods that do not result in complete and permanent changes to model parameters can still facilitate the extraction of sensitive information. 
%JOSEP. Moved here
The study by \cite{maini2024tofu} evaluates the effectiveness of several existing unlearning methods for LLMs. 
It concludes that none of these methods fully achieve effective unlearning, which indicates a need for developing more effective approaches that ensure models behave as if they were never trained on the data intended to be forgotten.
Large models contain many parameters representing numerous data dimensions and correlations, and this makes it hard to pinpoint the effect of specific training data.
Existing LLM unlearning methods are usually context and task-dependent and their generality remains unclear.

%JOSEP. Moved here.
{\em Potential solution.} The ideal method should focus on directly erasing sensitive data from model parameters, and ensuring that they remain inaccessible, thus complying with privacy regulations \citep{zhang2023right}. 
Additionally, this approach protects against the potential threat of extracting sensitive data through white-box attacks, particularly when LLMs are publicly available to adversaries possessing the technical expertise required to access sensitive information stored within model parameters or hidden states~\citep{patil2023can}.

%JOSEP. Added 
We next examine the limitations of the forgetting achieved by several families of method.

As an example, \cite{xu2023machine} classify some of the surveyed forgetting mechanisms as exact or strong, but most of them are simple models such as linear regression, which can be amenable to such kinds of analyses.
On the other hand, given some of the approaches, like fine-tuning using adapters (including LoRA), the applicability of these guarantees seems challenging since they make obvious changes to the model's architecture. At most, such forgetting procedures could be considered as providing \textit{weak} forgetting guarantees.

% Challenges to RLFH
%JOSEP. A bit rewritten and moved here.
Another example are methods based on RLHF. These may still know the information they were supposed to forget, which could be triggered and produced with adversarial prompts \citep{zou2023universal}.

% Najeeb: Challenges of direct modification methods
%JOSEP. A bit rewritten.
Regarding direct modification methods, while they offer a promising approach to efficiently and permanently remove to-be-forgotten information from LLMs, \cite{patil2023can} observed that sensitive information that is `deleted' from LLMs by model editing can still be extracted from the model's hidden states. Also, applying an editing method for one question may not effectively erase information across rephrased versions of the same question. 
These shortcomings are due to the non-linearity of LLMs. 

% Najeeb: Challenges of Black-box methods. 
%JOSEP. Moved here.
Black-box methods do not completely erase the to-be-forgotten knowledge from the model because it does not alter the model's parameters. This might not be compliant with privacy and copyright regulations. 

% Najeeb: Challenges of in-context learning
%JOSEP. Moved here.
In-context learning methods may be prone to knowledge conflict issues \citep{zhang2024comprehensive} and their unlearning effectiveness depends on the in-context learning capability of the used LLMs.
A recent study by~\cite{yu2023characterizing} explores different scenarios where LLMs choose between in-context and memorized answers. 
These scenarios often involve a trade-off between relying on the model’s pre-existing knowledge and the new information provided in the context. 
This study underscores the need for further investigation into when and how to use in-context learning methods.

\subsection{Retaining of model utility}

% Najeeb: Challenges to loss maximization methods
%JOSEP. A bit rewritten and moved here.
Methods based on loss maximization are very sensitive to 
hyperparameter choosing ({\em e.g.}, learning rate, number of fine-tuning epochs).
A wrong choice of parameters can cause the model to lose its language understanding.

%JOSEP. Moved here
On the other hand, unlearning a single fact can have intricate implications due to the interconnected nature of knowledge acquired by LLMs. 
This interconnection means that the removal or modification of one piece of information can potentially ripple through the network, altering the way other facts are represented or understood.
Given this complexity, there is a need for a more profound investigation and comprehensive evaluation of the unlearning effects. 
Such an investigation would not only illuminate the direct consequences of unlearning but also reveal any indirect effects caused by the interconnectedness of knowledge.

% Najeeb. Requiring retain set for unlearning
%JOSEP. Moved here.
Also, the requirement of the retain set for unlearning is not only restrictive but also poses an unrealistic assumption \citep{maini2024tofu}. 
Therefore, it becomes crucial for contemporary methods to pivot towards a more feasible approach. 

%Najeeb: Regarding retainability
%JOSEP. Moved here and rewritten.
Another thorny issue is the model’s ability to maintain its utility in the face of numerous unlearning requests.
How does the frequency of unlearning requests influence the overall performance of the model? 
This is a crucial area of study to ensure the robustness of LLMs amidst continuous updates and modifications.
Existing methods predominantly utilize static forget rates for performance evaluation, and typically process forget requests in a single batch. 
This approach may not accurately represent real-world scenarios where forget requests may occur sporadically and at varying intervals. 

{\em Potential solution.} Future research should focus on examining individual forget requests that arise sequentially but at distant time intervals. 
This would provide a more comprehensive understanding of their impact on the model’s utility and scalability, especially when dealing with a large volume of successive forgetting requests. This shift in focus could potentially lead to more robust and adaptable LLMs.

\subsection{Generalization of unlearning}

%JOSEP. Moved here.
% Najeeb: No single method can effectively unlearn sensitive data
\cite{patil2023can} proposed six different defense methods to counter extraction attacks but concluded that no single universally effective defense method exists against all extraction attacks.

{\em Potential solution.} One could combine one or more unlearning methods with model editing to create a multifaceted and effective defense strategy against extraction attacks.

% Najeeb: No general method suitable for all purposes
In existing methods, there is no general method that is suitable for all forgetting/unlearning purposes. Some methods are suitable for meeting privacy/copyright requirements. Other methods are suitable for debiasing models. 

{\em Potential solution.} One could combine several specialized methods to obtain a general unlearning method.

%JOSEP. timeliness -> runtime
\subsection{Runtime and scalability}

%JOSEP. Moved here.
%Najeeb. When unlearning is the suitable option
The unlearning approach is an appropriate choice when the priority is to stop generating undesirable outputs rather than trying to generate desirable ones.
It becomes more attractive when computational resources are limited or the model operator does not want to spend significant time eliminating undesirable outputs.

%%%%%%%%%% Taking stuff from below
%Najeeb: General challenges to white-box methods
%JOSEP. A bit rewritten.
White-box methods encounter several challenges. 
These include computational and memory requirements due to the huge number of parameters, risks of overfitting, and catastrophic forgetting. 
Such methods are widely used but: i) they are likely to overfit when data used for fine-tuning are small; ii) they incur high computational costs even with efficient fine-tuning methods; iii) they could lead to losing learned knowledge in the pre-trained parameters due to modifying them with no constraints.

% Najeeb: Challenges to training data sharding methods
%JOSEP. A bit rewritten.
Methods relying on training data sharding provide a stronger forgetting guarantee but require significant computational and storage resources, and they are time-consuming. With LLMs, they are not practical.
Moreover, these methods are unsuitable for addressing biases in LLMs. 
In instances where bias patterns permeate across all shards, necessitating retraining of all segments, such methods will require significant computations.

%JOSEP. Moved here
{\em Potential solution.} %Najeeb. The potential of local fine-tuning and direct local modification
Progress is being made in modular machine learning~\cite{feng2023cook}, a field that revolves around the concept of dividing fundamental models into distinct, manageable modules. 
Each module's specific knowledge can be then updated individually. 
This structure paves the way for local modification methods that focus on unlearning specific knowledge once it's been localized within an LLM. 
This makes these methods increasingly feasible and holds great promise for delivering a form of unlearning that is permanent, lightweight, and utility-preserving.

%JOSEP. New subsection.
\subsection{Evaluation}

% Najeeb: Dynamics of leaking
%JOSEP. I suppress this. Talking of data points means this is not on LLMs, right?
%\cite{borkar2023what} found that once the data points that are highly vulnerable to leakage are unlearned, a new set of data points that were previously safe become vulnerable to leakage.

%Benet: No common benchmark, nor dataset nor model for comparable measures
The literature on digital forgetting for LLMs exhibits a wide variety in terms of forgetting and retaining datasets, selected models, and evaluation metrics. 
This makes it impossible to fairly compare methods, and thereby ascertain the relative effectiveness of different approaches for specific forgetting requests. 
There is an evident need for a standard benchmark establishing a common testing ground for diverse unlearning methods. 
Such a benchmark should include a comprehensive and carefully selected set of forgetting requests (including forgetting and retaining datasets), models and metrics for forgetting, retaining, and speed.
% Benet NOTE: "For example, no pair of the methods analyzed can be fairly compared by leveraging the results provided in their papers." is false. For instance, in \cite{pawelczyk2023context} they compare their method with \cite{jang2022knowledge}. Moreover, there are some methods comparing themselves with previous unlearning approaches, but these previous techniques are not included in this survey.

%Benet: Difficult forgetting evaluation
Evaluating the success of forgetting poses a major challenge. There are multiple ways in which a model can be induced to produce results based on the undesired knowledge. 
%JOSEP. A bit rewritten.
Inputting the data leveraged for unlearning ({\em i.e.}, forgetting training set) or disjoint but from the same distribution ({\em i.e.}, forgetting test set) is easy. 
Nonetheless, \cite{eldan2023harry} show that undesired generations can be also obtained from out-of-distribution prompts ({\em e.g.}, asking to list fictional schools after ``unlearning'' the Harry Potter corpora and obtaining ``Hogwarts''). 
Moreover, the ``poem, poem, poem...'' case depicted in \cite{nasr2023scalable} indicates that even out-of-distribution inputs can cause the model to come back to the
%JOSEP. Added "to be forgotten" 
training text to be forgotten. 
Beyond the challenge of identifying ways in which the model preserves undesired knowledge, there lies the issue of determining how to effectively measure its actual preservation. 
For numerous forgetting requests, merely preventing the generation of verbatim duplicates of the forgetting training set is insufficient; any prediction indicating that the LLM retains that undesired knowledge needs to be avoided. For example, harmful or toxic generations can occur in a wide (even unlimited) variety of ways. Moreover, users actively define workarounds ({\em e.g.}, fine-grained paraphrasing and cheat prompts \footnote{\url{https://www.reddit.com/r/ChatGPT/comments/11yvk5g/using_the_neurosemantic_invertitis_hack_can/}}) to deceive toxicity detection mechanisms, thereby transforming the process into an ongoing adversarial challenge. This complicates the definition of a forgetting test set that is representative of potential error scenarios and of a metric to confidently assess whether 
%JOSEP. Slightly rewritten.
resulting predictions remain undesirable.
% Najeeb: Single dataset for evaluating a forgetting task
Almost all methods use a single dataset to evaluate the forgetting of a target task (privacy, copyright). 
To ensure the method is not dataset-specific, diverse datasets for the same target task must be used.
Unlearning in LLMs is more challenging than in traditional classification models due to the vast output space of language models (their outputs are not just a class label), higher efficiency requirements, and limited access to training data. This makes evaluations difficult.
%Alberto: qué tal así?
%Benet: Me parece bien, gracias. Lo he integrado en el texto y modificado algunas cosas que hablé con David
%In fact, difficulties in the detection of problematic concepts, such as harmful text generation, have become one of the main motivations for digital forgetting mechanisms. If model owners could identify when a generation is problematic, they could simply refrain from displaying it to the user. In contrast, digital forgetting renders useful to preclude problematic generations. For example, forgetting how a bomb is built to make it impossible to provide any useful information about it. However, forgetting about evaluation often requires these challenging detection metrics, defining an interdependency.
%Benet DOUBT: No estoy nada seguro de ^este^ último párrafo (el comentado). Sobretodo lo de "detection of problematic concepts, such as harmful text generation, have become one of the main motivations for digital forgetting". Si consideras mejor borrarlo, lo entendería

% Scalability: some methods may be okay to unlearn 1 or a few concepts, but what if we're requested to forget 100/1000s of concepts?

%Najeeb: Regarding evaluation

%Najeeb: Datasets
%JOSEP. uniform -> de facto standard
Currently, no {\em de facto} standard datasets exist for explicitly evaluating unlearning methods. 
Researchers typically assess these methods across diverse datasets based on individual considerations. 
Another challenge is that all methods are evaluated on English language datasets and there is no evidence about their performance with other languages.
%JOSEP. Rewritten.
More datasets on different unlearning applications are required to assess how well unlearning methods generalize across different languages. 

%JOSEP. Rewritten.
Furthermore, even if unlearning is motivated by concerns about private information leakage, the datasets used in the literature rarely contain sensitive information. 
To better validate the effectiveness of unlearning for sensitive information, future research should use simulated data sets with representative distributions of sensitive information.

%JOSEP. I delete this, because it has been said above, when talking about the runtime of sharding methods.
%Several unlearning methods heavily rely on retraining models to assess changes, which can be costly in the case of large language models.

%JOSEP. I deleted this, because this problem with bias 
%has already been commented above.
%Najeeb: Regarding bias evaluation
%Evaluating the language model's bias implicitly poses interpretability challenges. 
%It is not always straightforward, as biases can be implicit and subtly woven into the fabric of model responses.
%\cite{yu2023unlearning} suggest that discrete metrics may fail to capture nuanced biases, highlighting the need for improved evaluation metrics to assess the efficacy of debiasing methods.
%These improved metrics would provide a more accurate assessment of the effectiveness of debiasing methods, thereby ensuring the fairness and impartiality of LLMs.

%Najeeb: Evaluation benchmark
%JOSEP. I delete this. because it has already been said.
%Lack of a comprehensive evaluation benchmark of unlearning methods. TOFU \citep{maini2024tofu} tried to ...

%JOSEP. I delete this because bias has already been covered above.
%Najeeb. Bias mitigation methods challenges
%Measuring stereotypical assumptions in an open-ended setting is challenging.
%Automating the detection of stereotypical assumptions in free text is an open question for future research.

%Najeeb. When can each method be used
\subsection{When can each method be used?}

Despite the wide range of LLM unlearning methods, the choice among them depends on different factors and contexts. 

\begin{itemize}
\item Short-term hot fixes: Input/output modification methods might be the most suitable choice. 
These methods are typically quick to implement and do not require extensive knowledge about the model's internal workings.
\item Medium-term fixes: Local modification or fine-tuning-based approaches could be more appropriate. 
These methods offer a balance between speed and thoroughness, allowing for more targeted adjustments to the model.
\item Long-term fixes: Retraining the model from scratch while excluding the problematic data may be the best option. 
This approach is often accompanied by the availability of new training data and the obsolescence of previous data.
\end{itemize}

The required unlearning guarantee related to the target application also plays a crucial role in determining the method used. 

\begin{itemize}
\item To prevent bias, toxic predictions, or the production of copyrighted data, any method that can fix the issue (even a black-box method) will suffice. 
The primary concern here is ensuring the final output delivered to the end user does not contain undesirable behavior.
\item However, when it comes to privacy and fulfilling the right to be forgotten, a method is needed that ensures the associated knowledge is removed from the internal weights themselves and that provides the greatest forgetting guarantee possible.
\end{itemize}

It is important to note that most bias removal methods focus on identifying the source (concept) of bias in the model's internal representations. 
These methods may not be as useful in other applications such as privacy, copyright, and detoxification. 
Similarly, methods that focus on unlearning knowledge associated with the input tokens themselves may not be effective for bias reduction, as they do not address biased concepts produced later in the model's internal representation.

%Najeeb. When an LLM is used by a third party service via black-box access
\subsection{Black-box access scenario}
In recent times, there has been a surge in the utilization of LLMs across various services provided by third-party entities to the end users. In these scenarios, both the third-party service provider and the end user interact with the model as a black box, having no direct access to its internal workings or parameters.
Occasionally, there may arise a need to `forget' or eliminate certain undesirable behaviors that manifest in the model's output. This necessitates the intermediary operator, who mediates the interaction between the end user and the model, to have effective strategies in place to address such instances.

One promising approach to tackle this issue is the use of input/output modification methods. 
However, while these methods can be suitable for some cases, they may not always provide the optimal solution. 
A potentially more effective strategy could involve fostering increased interaction between the end user, the intermediary operator, and the primary operator of the model, who has full access to its components. 
This collaborative approach could facilitate the development of more robust unlearning solutions, allowing for the dynamic adjustment of the model's behavior based on user feedback and real-time requirements.

%Najeeb. Reconciling effectiveness, utility and efficiency
\subsection{Reconciling effectiveness, utility and efficiency}
The primary challenge of unlearning in large language models (LLMs) is achieving a balance between three critical aspects: the effectiveness of unlearning, the utility of the unlearned model, and the efficiency of the unlearning process. As of now, no existing method has been able to satisfactorily reconcile these conflicting objectives.

%JOSEP. Rewritten.
For instance, a retraining approach such as SISA accomplishes exact unlearning. However, this exact guarantee comes at a significant cost. The retraining approach is computationally intensive and imposes significant memory demands, which makes it unsuitable for LLMs due to their large scale and complexity.

On the other hand, black-box methods, which operate without needing access to the model's internal parameters, are able to maintain the utility of the model and are generally more lightweight. However, these methods offer weak unlearning guarantees: they may not always effectively remove the undesired behavior from the model's output.

This challenge underscores the urgent need for future research to focus on finding a better alignment between these contradictory requirements. The goal should be to develop unlearning methods that are not only effective and efficient but also preserve the utility of the model. This would represent a significant advancement in the field, and it would enable safer and more flexible use of LLMs in a wide range of applications.

%JOSEP. Added section.
\section{Conclusions}
\label{sec:conclusions}

This document has surveyed the state of the art on unlearning in LLMs. We have first given background on LLMs. Then we have reviewed the motivation, the types, and the requirements of digital forgetting. Next, we have described the main approaches used by digital forgetting methods in LLMs. After that, we have surveyed the literature by grouping unlearning methods in four categories: global weight modification, local weight modification, architecture modification, and input/output modification. After the description of the literature, we have described the way the proposed methods are evaluated: which datasets are used; to which LLMs models is unlearning applied; the metrics and attacks used to measure forgetting; the metrics used to measure retaining on the tasks to be preserved after forgetting; and the runtime of unlearning methods. 

Last but not least, we have identified a good number of challenges in the current state of the art. It can be concluded that, if machine unlearning in general is a hot topic far from having reached maturity, this is even truer when we talk about machine unlearning in LLMs. The size and the unprecedented power of LLMs can only add challenges to this still nascent area.

%JOSEP. Added.
\section{Acknowledgments}

This work has been funded by Huawei Technologies Finland Research Center.

\addcontentsline{toc}{section}{References}
%David: changed to APA-like reference style to make more emphasis on the surveyed papers
\bibliographystyle{apalike}
\bibliography{references}

\end{document}